\documentclass[submitting]{nst}

\usepackage{subfigure,dcolumn}
\usepackage[T2A,T1]{fontenc}
\usepackage[russian,english]{babel}

\usepackage{listings}
\usepackage{xcolor}

\begin{document}

\title{A Silicon Microstrip Detector for Power-Limited and Large Sensitive Area Applications}\thanks{This study was supported by National Key Programme for S\&T Research and Development (Grant NO.: 2022YFA1604800), and the National Natural Science Foundation of China (Grant NO.: 12342503).}

\author{Dexing Miao}
\affiliation{Institute of High Energy Physics, Beijing 100049, China}
\affiliation{University of Chinese Academy of Sciences, Beijing 100049, China}
\author{Zijun Xu}
\affiliation{Institute of High Energy Physics, Beijing 100049, China}
\author{Zhiyu Xiang}
\affiliation{Institute of High Energy Physics, Beijing 100049, China}
\author{Pingcheng Liu}
\affiliation{Shandong Institute of Advanced Technology, Jinan
250100, China}

\author{Giovanni Ambrosi}
\affiliation{INFN Sezione di Perugia, 06100 Perugia, Italy}
\author{Mattia Barbanera}
\affiliation{INFN Sezione di Perugia, 06100 Perugia, Italy}
\author{Mengke Cai}
\affiliation{Institute of High Energy Physics, Beijing 100049, China}
\author{Xudong Cai}
\affiliation{Massachusetts Institute of Technology (MIT), Cambridge, Massachusetts 02139, USA}
\author{Hsin-Yi Chou}
\affiliation{Institute of Physics, Academia Sinica, Nankang, Taipei, 11529, Taiwan}
\author{Matteo Duranti}
\affiliation{INFN Sezione di Perugia, 06100 Perugia, Italy}
\author{Valerio Formato}
\affiliation{INFN Sezione di Roma Tor Vergata, 00133 Roma, Italy}
\author{Maria Ionica}
\affiliation{INFN Sezione di Perugia, 06100 Perugia, Italy}
\author{Yaozu Jiang}
\affiliation{INFN Sezione di Perugia, 06100 Perugia, Italy}
\author{Liangchenglong Jin}
\affiliation{Institute of High Energy Physics, Beijing 100049, China}
\author{Vladimir Koutsenko}
\affiliation{Massachusetts Institute of Technology (MIT), Cambridge, Massachusetts 02139, USA}
\author{Qinze Li}
\affiliation{Institute of High Energy Physics, Beijing 100049, China}
\affiliation{University of Chinese Academy of Sciences, Beijing 100049, China}
\author{Cong Liu}
\affiliation{Shandong Institute of Advanced Technology, Jinan
250100, China}
\author{Xingjian Lv}
\affiliation{Institute of High Energy Physics, Beijing 100049, China}
\affiliation{University of Chinese Academy of Sciences, Beijing 100049, China}
\author{Alberto Oliva}
\affiliation{INFN Sezione di Bologna, 40126 Bologna, Italy}
\author{Wenxi Peng}
\affiliation{Institute of High Energy Physics, Beijing 100049, China}
\author{Rui Qiao}
\affiliation{Institute of High Energy Physics, Beijing 100049, China}
\author{Gianluigi Silvestre}
\affiliation{INFN Sezione di Perugia, 06100 Perugia, Italy}
\author{Zibing Wu}
\affiliation{Shandong University (SDU), Jinan, Shandong 250100, China}
\author{Xuhao Yuan}
\affiliation{Institute of High Energy Physics, Beijing 100049, China}
\author{Hongyu Zhang}
\affiliation{Key Laboratory of Nuclear Physics and Ion-beam Application (MOE) and Institute of Modern Physics,
Fudan University, 220 Handan Road, Shanghai, 200433, China}
\author{Xiyuan Zhang}
\affiliation{Institute of High Energy Physics, Beijing 100049, China}
\author{Jianchun Wang}
\email[Jianchun Wang, ]{jwang@ihep.ac.cn}
\affiliation{Institute of High Energy Physics, Beijing 100049, China}

\begin{abstract}
A silicon microstrip detector (SSD) has been developed to have state of the art spatial resolution and a large sensitive area under stringent power constraints. The design incorporates three floating strips with their bias resistors inserted between two aluminum readout strips. Beam test measurements with the single sensor confirmed that this configuration achieves a total detection efficiency of $99.8 \, \%$ and spatial resolution $7.6 \, \mathrm{\mu m}$ for MIPs. A double-$\eta$ algorithm was developed to optimize hit position reconstruction for this SSD. The design can be adapted for large area silicon detectors.
\end{abstract}


\maketitle

\section{Introduction}
Silicon microstrip detectors (SSD) \cite{Seidel:2019hty,Wei:2020lpl} are extensively used for tracking purposes across 
several experiments in particle physics~\cite{LHCb:2008vvz,CMS:2008xjf,ATLAS:2008xda}, nuclear physics~\cite{He:2023svm}, as well as in many space-borne experiments ~\cite{PAMELA:2004seq,Bonechi:2007dk,AGILE:2011aa,Bulgarelli:2010zz,Sgro:2007zz,Cecchi:2008zza,Azzarello:2016trx,Dong:2015qma,Vievering:2020wqk} including Alpha Magnetic Spectrometer (AMS-02)~\cite{Lubelsmeyer:2011zz,Duranti:2013qfz}. Due to stringent constraints on power consumption in a space-borne experiment, the number of available readout electronic channels for SSD is limited, which necessitates careful optimization of detector designs.

\begin{figure*}[!htb]
\centering
\includegraphics[width=0.8\hsize]{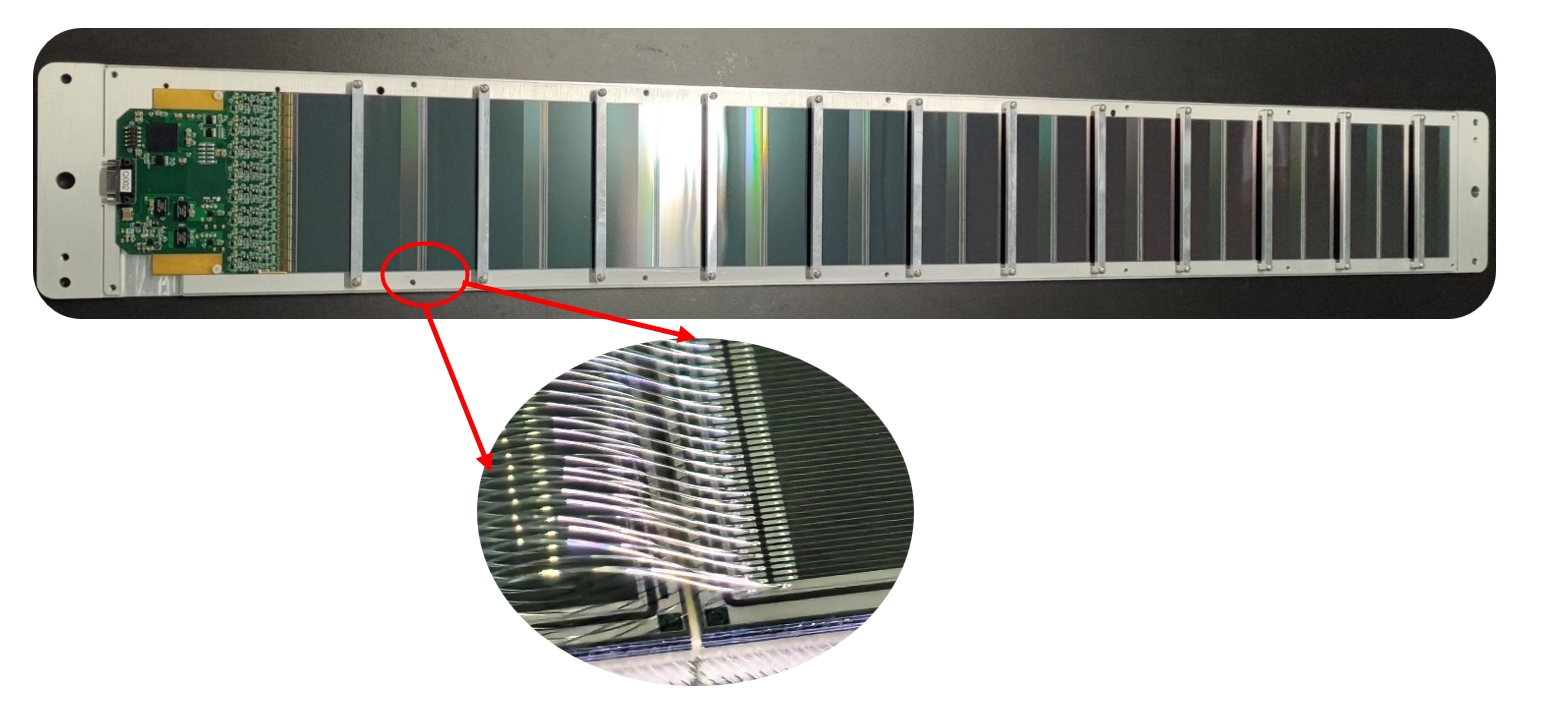}\\
\caption{A 12-SSD ladder of AMS Layer-0 tracker. This ladder is composed of 12 sensors connected through bonding wires (shown in the magnified figure).\label{fig:ladder}}
\end{figure*}

The AMS-02 experiment, operating aboard the International Space Station (ISS), aims to precisely measure the composition and flux of cosmic rays throughout the ISS's lifetime until 2030. To enhance detection acceptance by a factor of three, a new silicon tracker layer (Layer-0) will be stacked at the top of AMS-02 in 2026. Layer-0 deploys the same module design as AMS-02 flying hardware, which consists of long silicon ladders, each constructed by daisy-chaining multiple SSDs. In the Layer-0 upgrade modules, this concept will be even further stressed by daisy-chaining 8, 10, or 12 SSDs, for a total length of $65 - 100\, \mathrm{cm}$. As depicted in Fig~\ref{fig:ladder}, the 12-SSD ladder represents the longest silicon strip detector ever built, presenting unique technical requirements. Consequently, we have designed and implemented an SSD specifically tailored to meet these stringent requirements for all ladders in the AMS Layer-0 tracker.

In this paper, we present the detailed design of this SSD, its characterization through test beam, and performance evaluations focused on detection efficiency and spatial resolution.

\section{Designs of the Silicon microstrip detector}\label{sec:SSD}
\begin{figure*}[!htb]
\includegraphics
  [width=0.8\hsize]
  {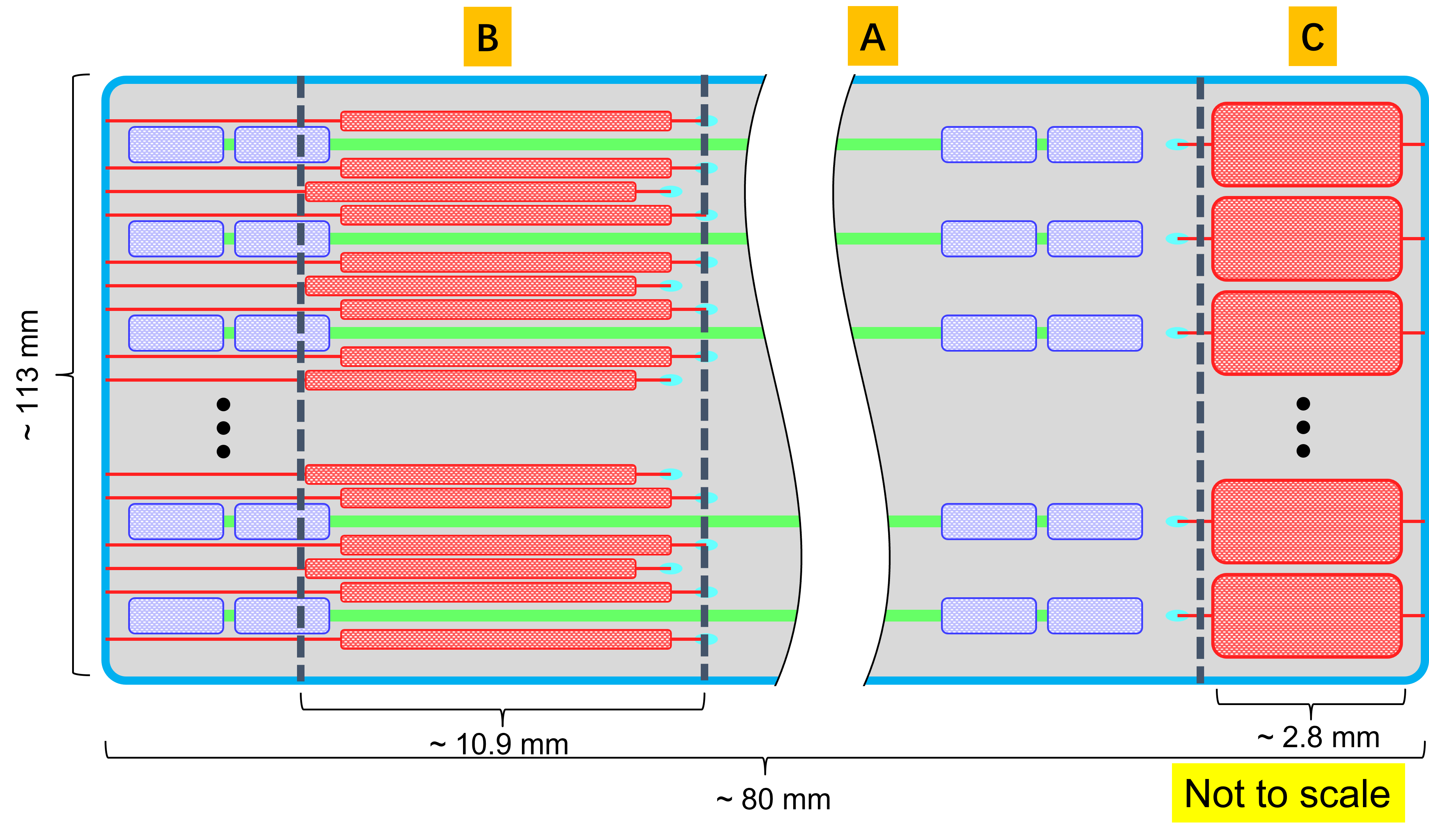}
\caption{The schematic diagram of the SSD design, where red boxes indicate bias resistors, blue boxes represent bonding pads, green strips denote readout aluminum strips. Different regions are separated by gray dashed lines and referred to as: (A) with three floating strips and without bias resistors; (B) with bias resistors for three floating strips located between two aluminum readout strips; (C) without floating strips and with bias resistors for readout strips. \label{fig:sensor}} 
\end{figure*}

The SSD is a $\rm{p}^+\mathrm{-in-n}$ sensor manufactured by Hamamatsu Photonics K.K, with a total area of $113 \times 80 \, \mathrm{mm^2}$ and thickness $320 \, \mu \mathrm{m}$, whose schematic is presented in Fig.~\ref{fig:sensor}. There are two specific designs: the use of three floating strips to enhance spatial resolution without increasing the number of electronic channels, and a configuration of large bias resistors and aluminum readout strips to ensure a large sensitive area. These designs result in different regions of the SSD, as indicated in Fig.~\ref{fig:sensor}: 

\begin{itemize}
\item Region-A) with three floating $\rm{p}^{+}$ strips, without bias resistors;
\item Region-B) with three floating $\rm{p}^{+}$ strips and their bias resistors located between two aluminum readout strips;
\item Region-C) without floating $\rm{p}^{+}$ strips, with bias resistors for $\rm{p}^{+}$ readout strips.
\end{itemize}

A $\rm{p}^{+}$ strip AC-coupled to an aluminum readout strip is referred to as the readout strip, while the strips located between two readout strips are called floating strips. In the SSD, which consists of 4095 $\rm{p}^{+}$ strips with a pitch of $27.25 \, \mathrm{\mu m}$, and one out of every four $\rm{p}^{+}$ strips is readout strip. There are three floating strips between adjacent readout strips. These floating strips facilitate charge sharing to readout strips, thereby enhancing spatial resolution~\cite{Turchetta:1993vu}. The impact of floating strips is quantitatively studied by comparing the performance between Region-C and Region-A.

In a ladder, aluminum readout strips of the \(N\) (\(N = 8, 10, 12\)) SSDs are connected in series along the strip direction, forming a parallel connection between the SSDs. Consequently, the coupling capacitors and bias resistors are connected in parallel. This arrangement leads to the coupling capacitance (bias resistance) increasing (decreasing)  by a factor of \(N\). Since the coupling capacitance cannot easily be adjusted during SSD fabrication by HPK, the SSDs must incorporate a bias resistance that is \(N^2 \sim 100\) times larger than the typical value in order to proper impedance matching.

For a conventional SSD, the bias resistance is $\mathcal{O}(1)\, \mathrm{M}\Omega$, implemented using polysilicon resistor that occupies approximately $100 \, \mu\mathrm{m}$ in length at the end of the aluminum readout strip~\cite{Casse:2013xia,lhcb_velo,CMS_SSD_bias_resistor,itk_bias_resistor}. In the SSD design, the $\mathcal{O}(100) \, \mathrm{M}\Omega$ bias resistor would occupy length of $\mathcal{O}(10) \, \mathrm{mm}$. Placing it directly behind the aluminum readout strips would lead to a bonding wire of about $14 \, \mathrm{mm}$ in length, which poses a potential risk of short-circuit. Therefore, a specialized bias resistor configuration was designed, as illustrated in Fig.~\ref{fig:sensor}. The bias resistors for the three floating strips are inserted between two aluminum readout strips, allowing the bonding pads to be placed at the left end of the aluminum strips. On the right end, the bias resistors of the readout strips occupy that space solely, requiring only $2.9 \, \mathrm{mm}$ in length. Thus, the length of bonding wires between two SSDs is reduced to about $5\, \mathrm{mm}$. 

The bias resistors between two aluminum readout strips would potentially influence their AC coupling or distort the internal electric field. Therefore, we carried out studies to compare the performance between regions with (Region-B) and without (Region-A) the bias resistors.

\section{Beam test setup}

The beam test described in this work was conducted in May 2024 at the Super Proton Synchrotron (SPS) at CERN. We performed the beam test behind the dumper of beam line H6. The beam mostly consisted of muons with momentum in the tens of GeV, generated by the proton beam striking the dumper, thus behaving as minimum ionizing particles (MIPs).

\begin{figure}[!htb]
   \subfigure[]{
   \begin{minipage}[t]{0.45\hsize}
   \centering
   \includegraphics[width=\hsize]{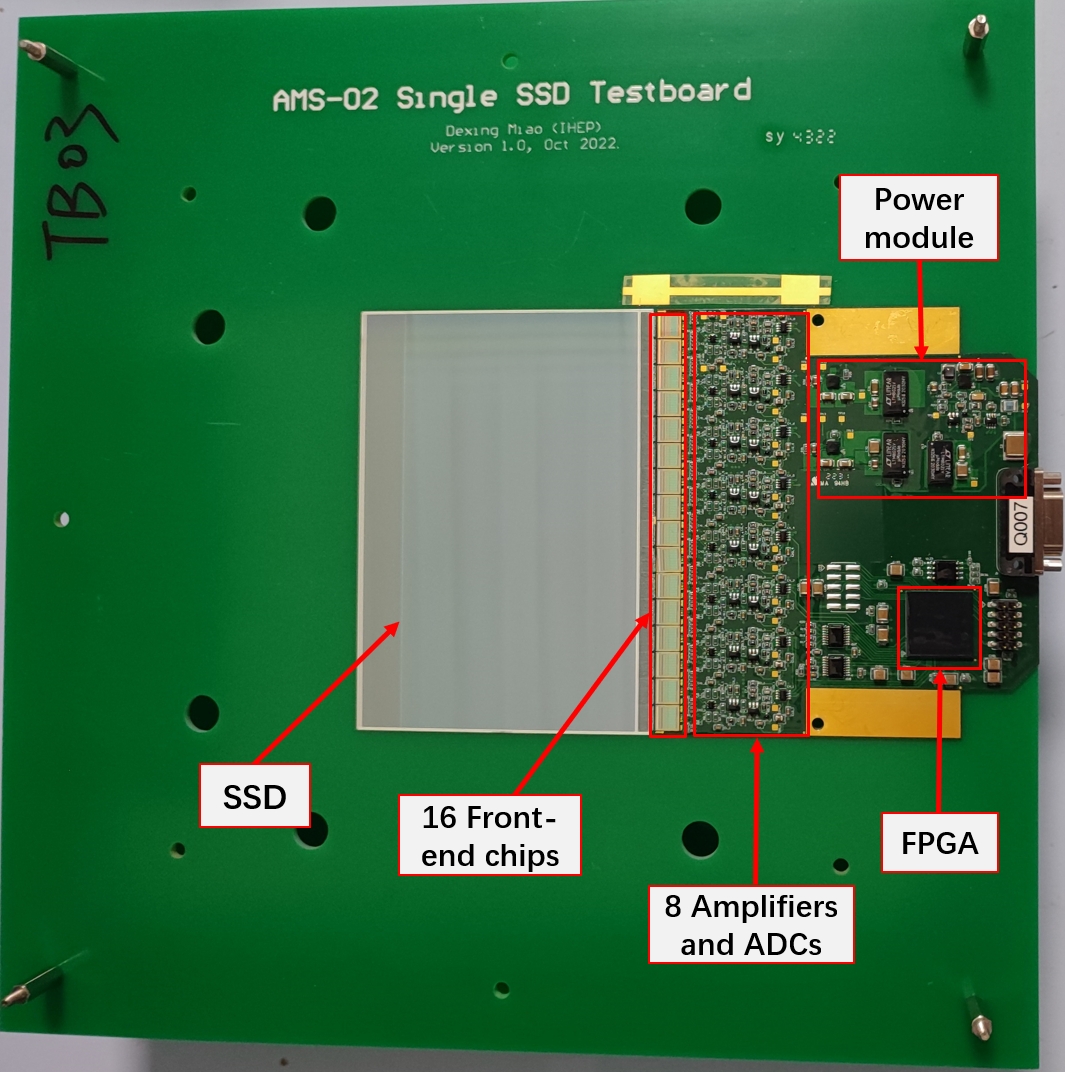}
    \label{fig:SSD_frame_board}
    \end{minipage}
    }
    \subfigure[]{
    \begin{minipage}[t]{0.45\hsize}
    \centering
    \includegraphics[width=0.8\hsize]{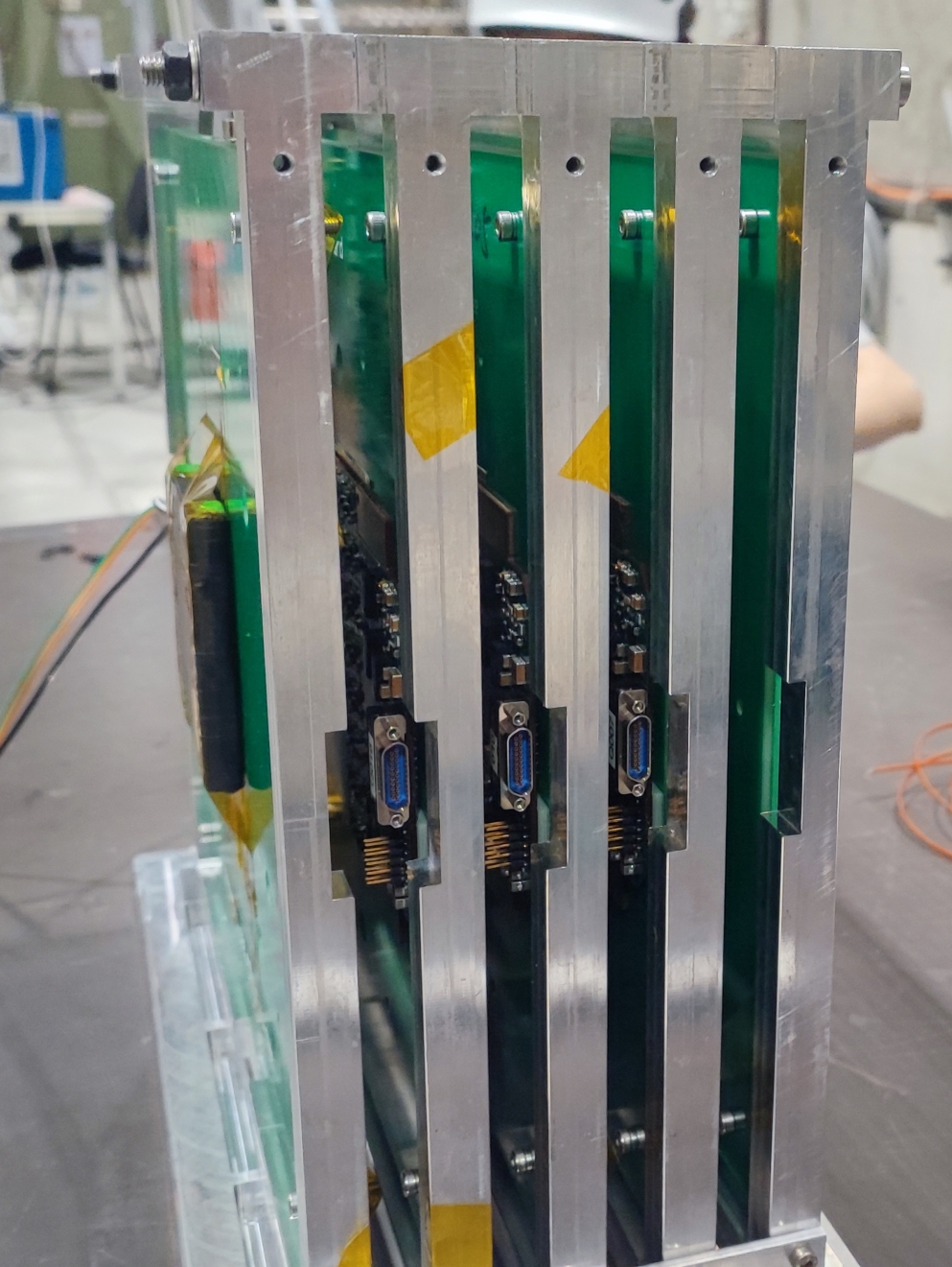}
    \label{fig:beam_monitor}
    \end{minipage}
    }
\caption{(a) Photograph of single SSD frame board, using the same electronics as the Layer-0 ladder, together with a description of each component. (b) Photograph of a beam monitor with several single SSD frame boards mounted on the aluminum frame.}
\label{fig:subfigures}
\end{figure}

We designed the single SSD frame board as shown in Fig.~\ref{fig:SSD_frame_board}. The gray rectangle in the middle is the SSD, which is wire-bonded to 16 front-end readout chips (IDE1140)~\cite{IDE1140} on the right, resulting in a total of 1024 channels. Each channel has a gain of $2.6 \, \mu \mathrm{A/fC}$ and a peaking time for positive signals of $8.5 \, \mu \mathrm{s}$. The output signals from IDE1140 are amplified in two stages with a gain of $6.97 \, \mu \mathrm{A/fC}$ and then routed to a 14-bit analog-to-digital converter (ADC) with a maximum input of $4096 \, \mathrm{mV}$. The data are then encoded by a field-programmable gate array (FPGA) and transmitted to the back-end.

\begin{figure}[!htb]
\centering
\includegraphics[width=\hsize]{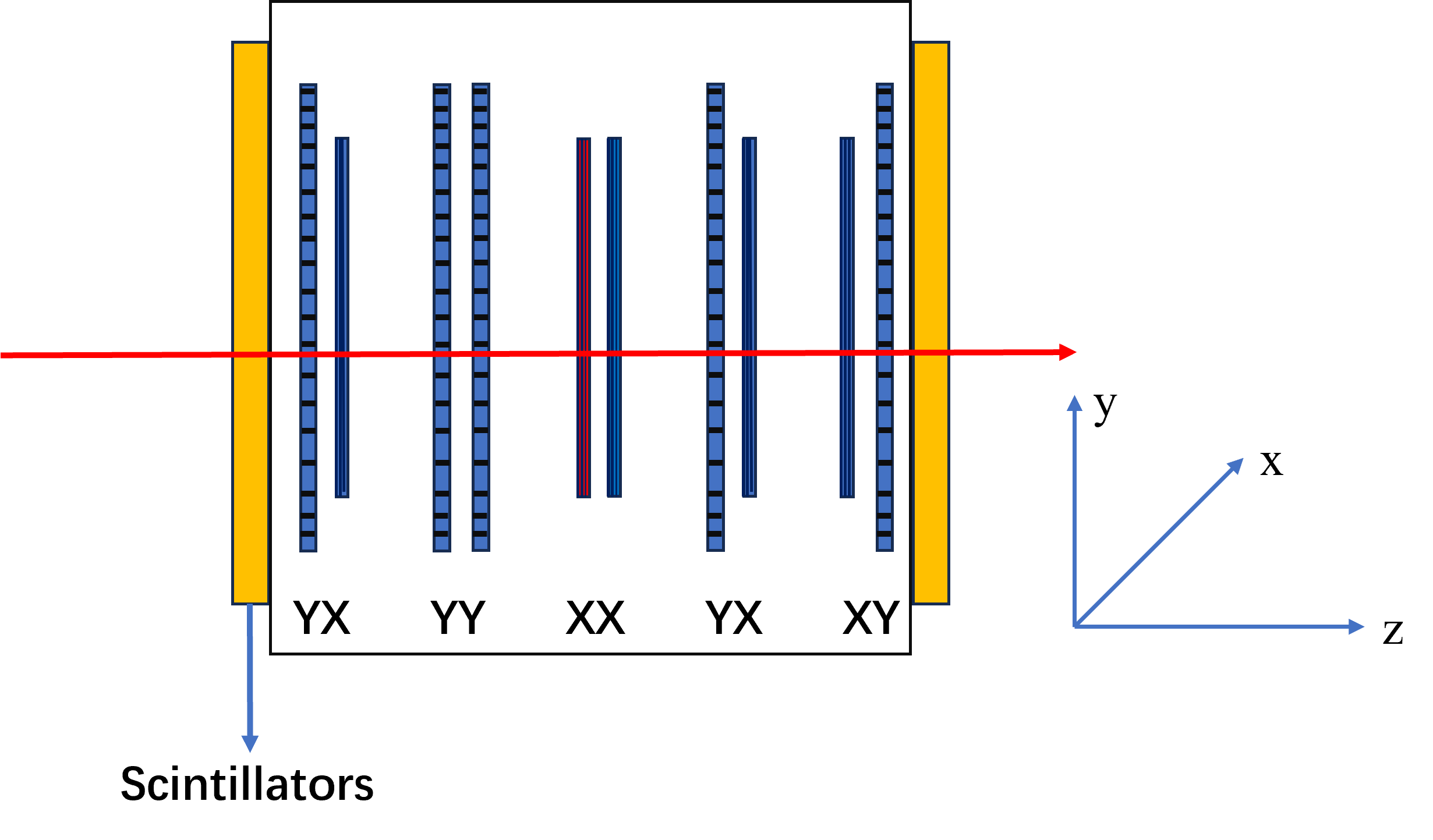}
\caption{Schematics of the layout of the beam test setup. Blue layers form the telescope. The red layer is DUT. Label X represents the measurement of the X coordinate, and Y likewise. Scintillator detectors are placed in overlapping pairs at both ends to provide the trigger signal.
\label{fig:effi_setup}}
\end{figure}

The beam monitor was constructed by several layers of single SSD frame boards, as depicted in the right panel of Fig.~\ref{fig:beam_monitor}. The frame board is designed with 90-degree rotational symmetry, allowing flexible orientation combinations for the layers. For layers with strips along the vertical (horizontal) direction, the hit position in the X (Y) coordinate can be measured, denoted as X-layer (Y-layer). In this beam test, we arranged 5 X-layers and 5 Y-layers, as illustrated in Fig. \ref{fig:effi_setup}, with the indicated detector under test (DUT) and the telescope. Scintillator detectors ($140 \times 80 \, \mathrm{mm^2}$)~\cite{Zhang:2023mti} were deployed to generate trigger signals, forming a trigger region that fully covered the DUT along the strip direction. This arrangement enabled performance evaluation of the DUT across Regions A, B, and C.

\section{Data Analysis}
% \qquad The data analysis starts with raw data and ultimately assesses the performance of the SSD. This process incorporates several design-specific features of our SSD, which are discussed in detail in the following sections.

\subsection{Raw data processing} \label{calibration}
The raw ADC value of each channel has four components:
    \begin{equation}
        {ADC}_{i}^{j} = {Ped}_i + {CN}_{i}^{j} + {Noise}_{i}^{j} + S_{i}^{j}\;,
    \end{equation}
where $i$ and $j$ are the channel number and event number, respectively. The ${Ped}_i$ is the pedestal of channel $i$, defined as the average value in the absence of particle incidence. Bias voltage fluctuation would cause an overall fluctuation of all channels, named common mode noise ($CN$)~\cite{Cui:2023ylq}. The single event $Noise_{i}^{j}$ is a random noise resulting from combining SSD and electronic noise, whose absolute value cannot be obtained, though the noise level of the channel $i$ ($Noise_i$) can be calculated as the standard deviation of ADC values after $CN$ is subtracted. The noise level of one channel is about 3 Least Significant Bits (LSB), which corresponds to Equivalent Noise Charge (ENC) $ \sim 240 \, e^-$. Finally, $S_i^j$ is the signal introduced by the incident particle.

The pedestal and noise level are determined through a one-minute calibration run conducted every hour during the beam test. For each event, $CN$ is calculated by:
    \begin{equation}
        CN^{j} =\frac{1}{N_a} \sum_{i = 1 }^{N_a}{ADC_i^j - Ped_i} \;,
    \end{equation}
where $N_a$ is the number of valid channels on the $a^{th}$ IDE1140. Noisy channels ($Noise_i > 10 \, \mathrm{LSB}$) and the fired channels ($ADC_i^j - Ped_i) > 10\, \mathrm{LSB}$) are not used in the $CN$ calculation.

Subsequently, for the data-taking run, calculate the signal of channel $i$ of event $j$: 
\begin{equation}
    S_{i}^{j} = ADC_{i}^{j} - Ped_i - CN_{i}^{j}\;.
\end{equation}
Then a cluster is identified by a channel where
\begin{equation}
    {S_i^j} > th_{\mathrm{seed}} \cdot Noise_i\; .
\end{equation}
The cluster expands on both sides until the value of channel \(x\) satisfies \(S_{x}^j < th_{side} \cdot {Noise}_{x}\).  In this study, $th_{\mathrm{seed}} = 3.5 $ and $th_{\mathrm{side}} = 2.0$ are chosen to reduce noise influence and maintain detection efficiency.  

\subsection{Double-$\eta$ position finding algorithm} \label{sec:doubel_eta}
For SSDs under normal incidence, the particle impact position is typically reconstructed from top few channels with the highest values in a cluster. Numerous studies have shown that the $\eta$ algorithm~\cite{Turchetta:1993vu,Abba:2015qka,Hubbeling:1991dx,Kraner:1983ky} is a proper position finding algorithm for this purpose. This algorithm reconstructs the particle impact position by using the two adjacent channels with the highest values in a cluster. Denote $S_{L}$ ($S_R$) be the value of the left (right) channel, $x_L$ ($x_R$) be the position of the readout strip corresponding to the left (right) channel. Define the variable $\eta$ :
\begin{equation}
\eta = \frac{S_R}{S_L + S_R} \;,\label{eq:eta}
\end{equation} 
The particle impact position $x_{\eta}$ follows a function of $\eta$:
\begin{equation}
x_{\eta} = x_L + Pf(\eta) \;, \label{eq:x_eta}
\end{equation}
where P is the readout strip pitch, $f(\eta)$ is the cumulative distribution function (CDF) of $\eta$:
\begin{equation}
    f(\eta)  = \int_{0}^{\eta}{\phi(\eta^{\prime}) d\eta^{\prime}} \;,
\end{equation}
$\phi(\eta)$ is the probability density function (PDF) of $\eta$, which can be estimated from test beam data. 

An issue arises when this algorithm is applied to the SSD, which exhibits significant charge sharing. For incidents near one readout strip, the two adjacent readout strips often share similar signals. Under the influence of random noise, the second-highest strip may be incorrectly identified. In these instances, choosing the 2 highest strips for position reconstruction can lead to a flip across the central strip, causing a larger position error. To address this issue, we proposed an iterative position finding method, namely ``double-\(\eta\) algorithm":
\begin{itemize}
    \item Picking the channel with the highest value in a cluster as the first channel. Among its two neighboring channels, the one with the larger value is denoted as the second channel, and the other as the third channel.
    \item Using first and second (third) channels to calculate two \(\eta\) values according to Eq.~\ref{eq:eta} respectively, denoted as \(\eta_{12}\) (\(\eta_{13}\)). Estimate two impact positions using Eq.~\ref{eq:x_eta}, denoted as $x_{{12}}$ ($x_{{13}}$).
    \item Using $x_{{12}}$ of all layers to reconstruct the initial trajectory.
    \item For each layer, compare the residual of $x_{12}$ and $x_{13}$ to the initial trajectory, select the smaller one as $x^{\star}$. Use $x^{\star}$ of all layers for final track reconstruction.
\end{itemize}

The spatial resolution comparison between this algorithm and the traditional $\eta$ algorithm has been discussed in Sec~\ref{sec:compare_double_eta}.

\subsection{Track Reconstruction}
Due to multiple particle incidences or electronic  noise fluctuations, a detector layer might have two or more clusters in a single event. In this beam test, approximately $10\%$ events contained two or more clusters. For these events, we evaluated all possible trajectory combinations and selected the one with the lowest $\chi^2$ value.

Track fitting is performed using the General Broken Lines (GBL) algorithm~\cite{Kleinwort:2012np}, widely employed in experiments such as AMS-02~\cite{Yan:2023xtc} and CMS~\cite{Otarid:2023anx}. GBL is capable of accurately accounting for multiple scattering effects. Additionally, GBL provides a comprehensive covariance matrix for all track parameters upon refitting, making it especially advantageous for calibration and alignment tasks when integrated with Millepede II~\cite{Blobel:2011az}, which is a global parameter optimization tool designed to efficiently handle up to approximately one hundred thousand global parameters.

Detector alignment is performed using $1 \, \%$ of the collected data. Subsequently, track fitting is carried out for all events that used telescope layers. Following this procedure, more than 10 million tracks with precise hit information were reconstructed for analyzing the performance of the DUT.

\section{Performance of the SSD}
We studied the impact of the SSD designs by comparing the performance across different regions, with a primary focus on charge collection and sharing performance, detection efficiency, and spatial resolution.

\subsection{Charge sharing and charge collection}

\begin{figure*}[!htb] 
\centering 
\includegraphics[width=0.8\hsize]{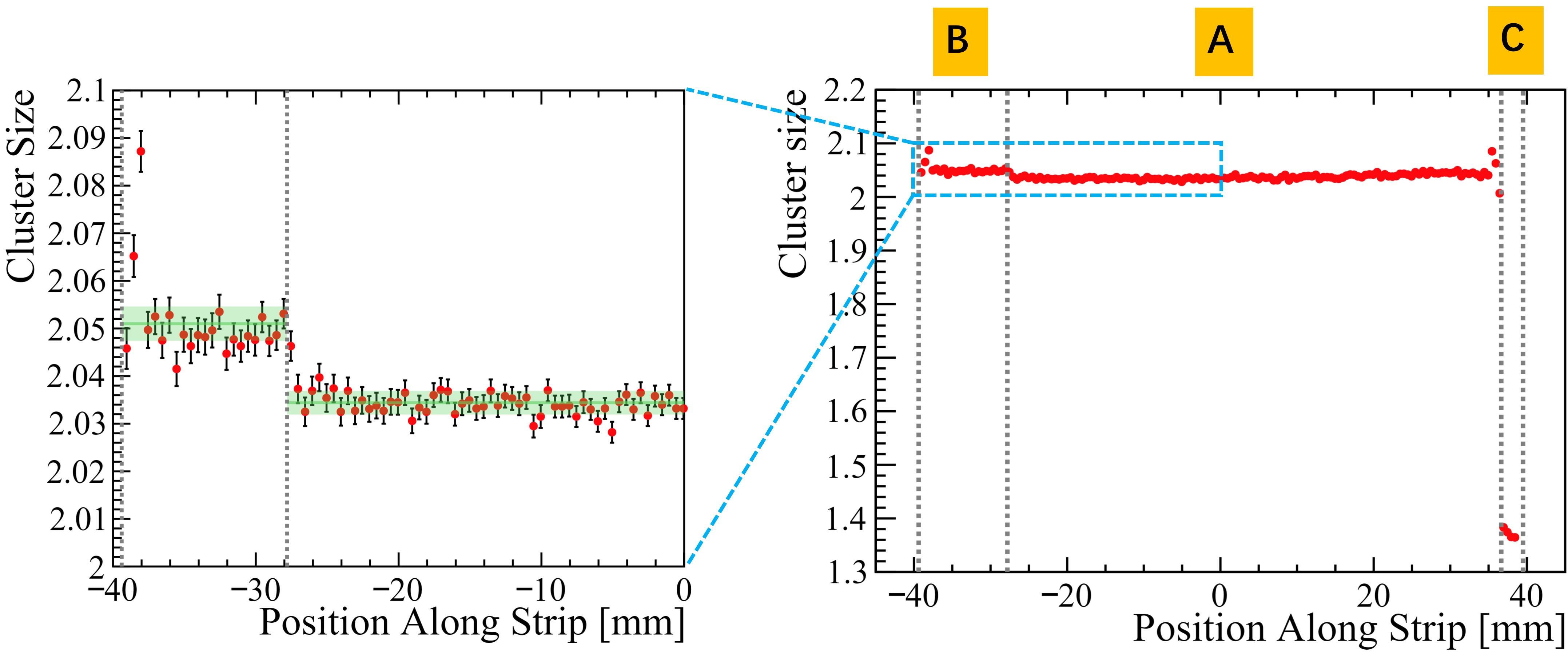}  
\caption{Right: the cluster size varies with the hit position along strip, with the dashed lines showing the SSD edge and the segmentation of the different regions as labeled. Left: magnified plot, where the green line and band indicate the mean cluster size and $\pm 1 \sigma$ error for Region-A and Region-B, respectively.\label{fig:cluster_size_along_strip}}
\end{figure*}

\begin{figure}[!htb]
   \subfigure[]{
   \begin{minipage}[t]{0.8\hsize}
   \centering
   \includegraphics[width=\hsize]{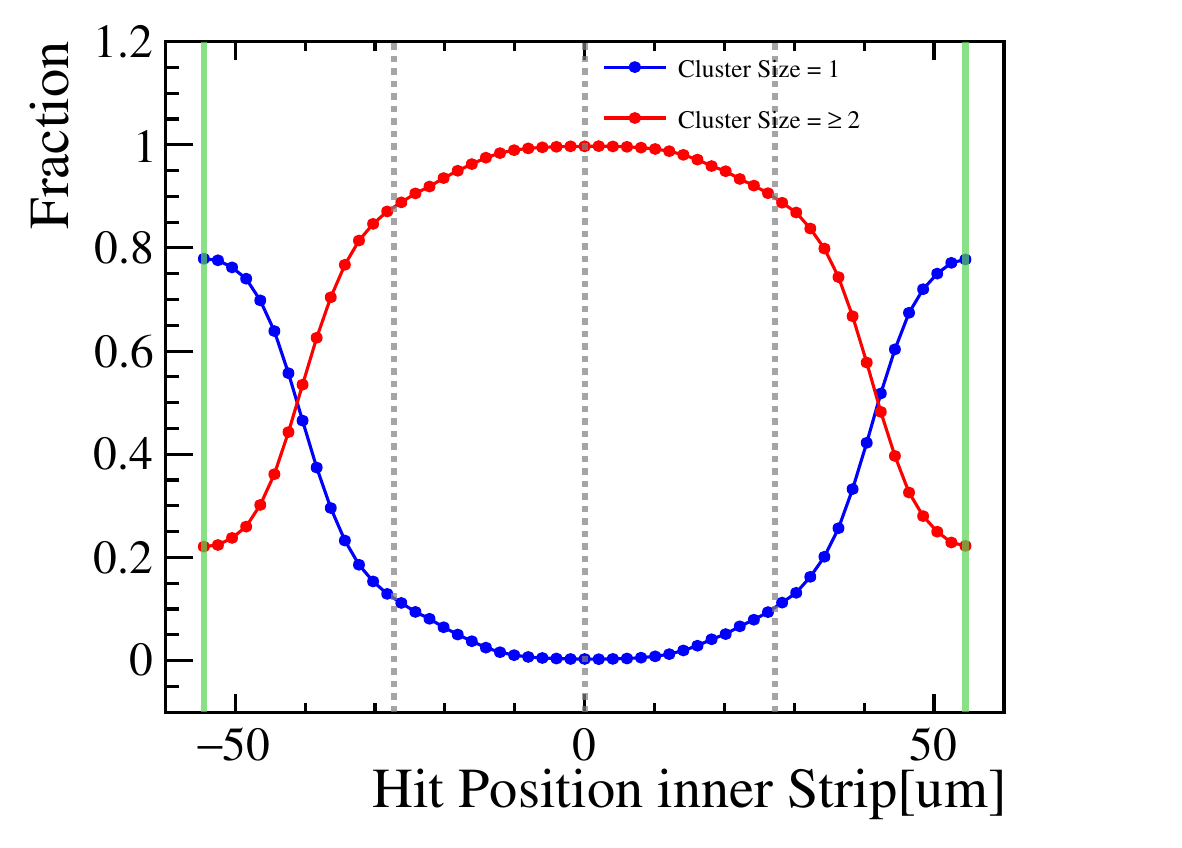}
    \label{fig:cluster_size_frac_A}
    \end{minipage}
    }
    \\
    \subfigure[]{
    \begin{minipage}[t]{0.8\hsize}
    \centering
    \includegraphics[width=\hsize]{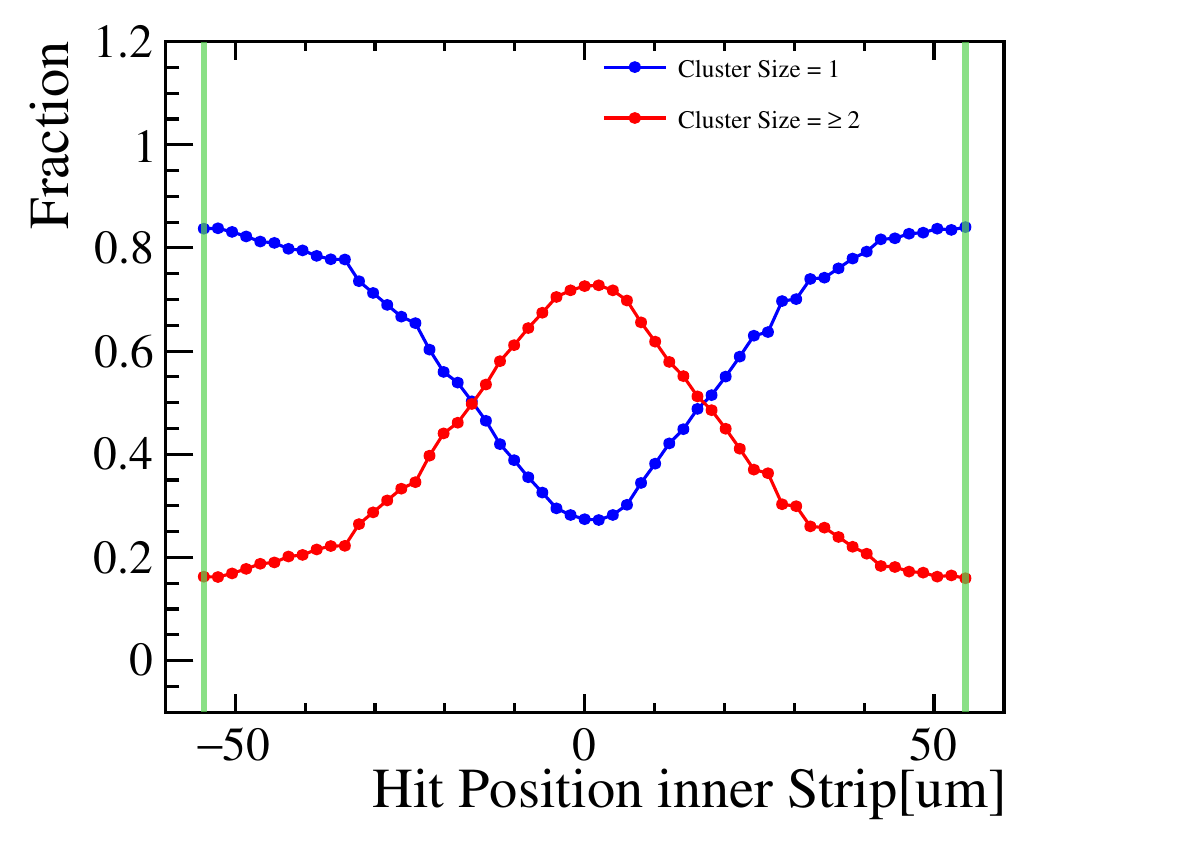}
    \label{fig:cluster_size_frac_C}
    \end{minipage}}
\caption{(a), (b) cluster size with inner strip hit position in Region-(A), Region-(C). The dashed line shows the position of the three floating strips, and the solid green lines represent the two readout strips.}\label{fig:cluster_size_frac}
\end{figure}

\begin{figure}
\centering
\includegraphics[width=\hsize]{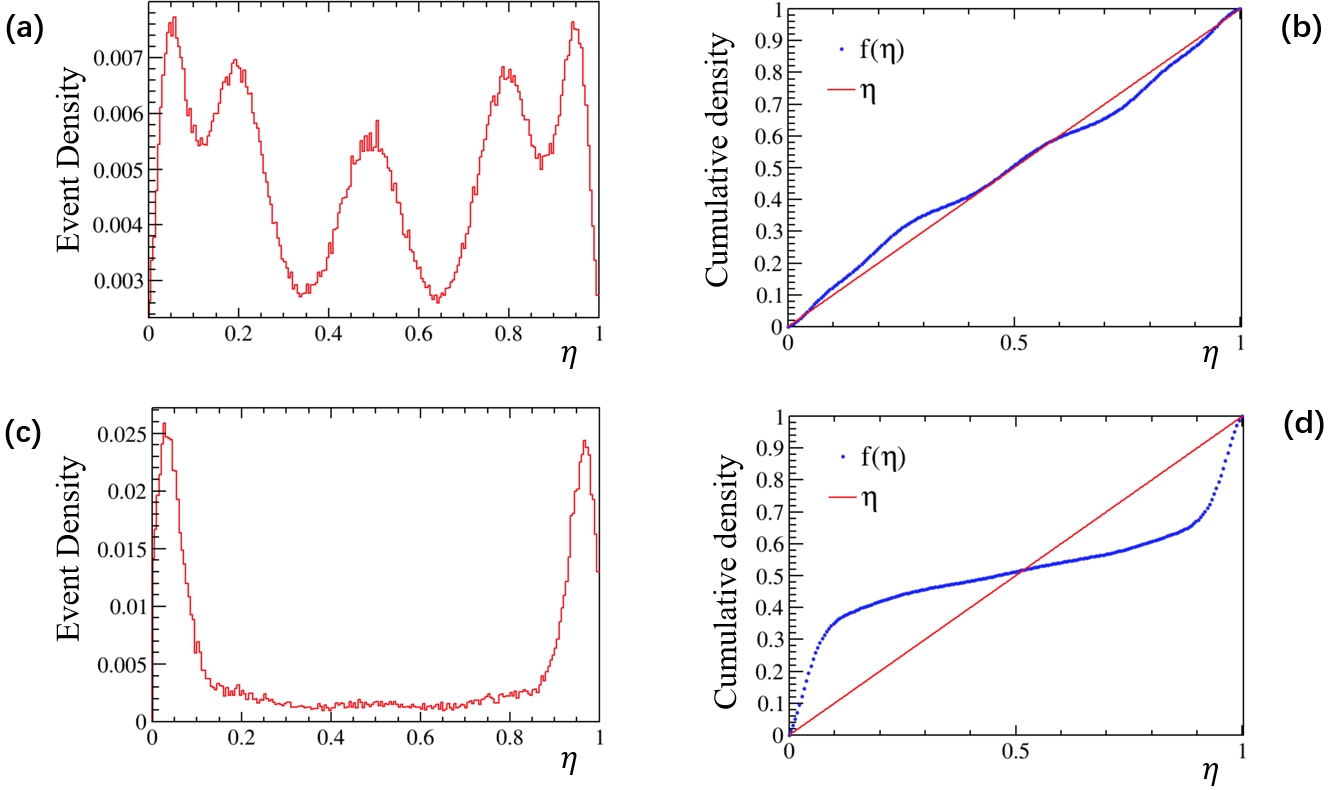}\\
\caption{The $\eta$ distributions of (a) Region-A and (c) Region-C, $f(\eta)$ of (b) Region-A and (d) Region-C. \label{fig:eta compare}}
\end{figure}

Charge sharing refers to the phenomenon where a single particle incident induces signals on multiple readout channels. The sharing is primarily due to capacitive coupling between neighboring strips and carrier diffusion. The charge sharing effect can be  quantified by the number of channels within a cluster, referred to as the cluster size. The right panel of Fig.~\ref{fig:cluster_size_along_strip} shows the variation of cluster size along the strip direction. The average cluster size is $2.03$ for Region-A and $1.38$ for Region-C, indicating the enhancement to charge sharing by the floating strips. Additional information can be gained by investigating the cluster size as a function of the relative position between two readout strips. For each track, the hit position inner strip is defined to be the position at the DUT mapped onto the range from [-54.5, 54.5], where -54.5 is the center of the $n^{\mathrm{th}}$ strip, and 54.5 is the center of the $(n+1)^{\mathrm{th}}$ strip. The fraction of the different cluster size as a function of hit position inner strip are shown in Fig.~\ref{fig:cluster_size_frac} for Region-A and Region-C. When floating strips are present, if the hit positions are slightly offset from the readout strips, there is a rapid increase in the fraction of clusters with a size $\, \geq 2$. Conversely, in areas without floating strips, clusters of size$\, \geq 2$ become dominant only when the hit occurs near the midpoint between two adjacent readout strips. This further underlines the role of floating strips in enhancing charge sharing.

The impact of the bias resistors for charge sharing located between two aluminum readout strips can be evaluated by comparing Region-B and Region-A. The magnified plot in Fig.~\ref{fig:cluster_size_along_strip} illustrates that the cluster size in Region-B is about $1\, \%$ larger than that in Region-A. This indicates that the presence of bias resistors leads to a few percent increase in charge sharing between aluminum readout strips, possibly due to the additional equivalent capacitance introduced by the polysilicon.

The variable $\eta$ also serves as an effective indicator of the degree of charge sharing. An $\eta$ value close to 0 or 1 indicates that most of the signal is collected by a single readout channel, while nearby 0.5 implies substantial sharing between two adjacent readout channels. Fig.~\ref{fig:eta compare}~(a)(c) shows a comparison of the \(\eta\) distributions between regions with and without three floating strips. It can be seen that in the region without floating strips, two prominent peaks appear near \(\eta = 0\) and \(\eta = 1\). In contrast, the presence of three floating strips introduces three additional peaks of \(\eta\) distribution. These peaks instruct that a specific fraction of charges is shared with the neighboring readout strip with the help of floating strips. 

\begin{figure}[!htb]
   \subfigure[]{
   \begin{minipage}[t]{0.45\hsize}
   \centering
   \includegraphics[width=\hsize]{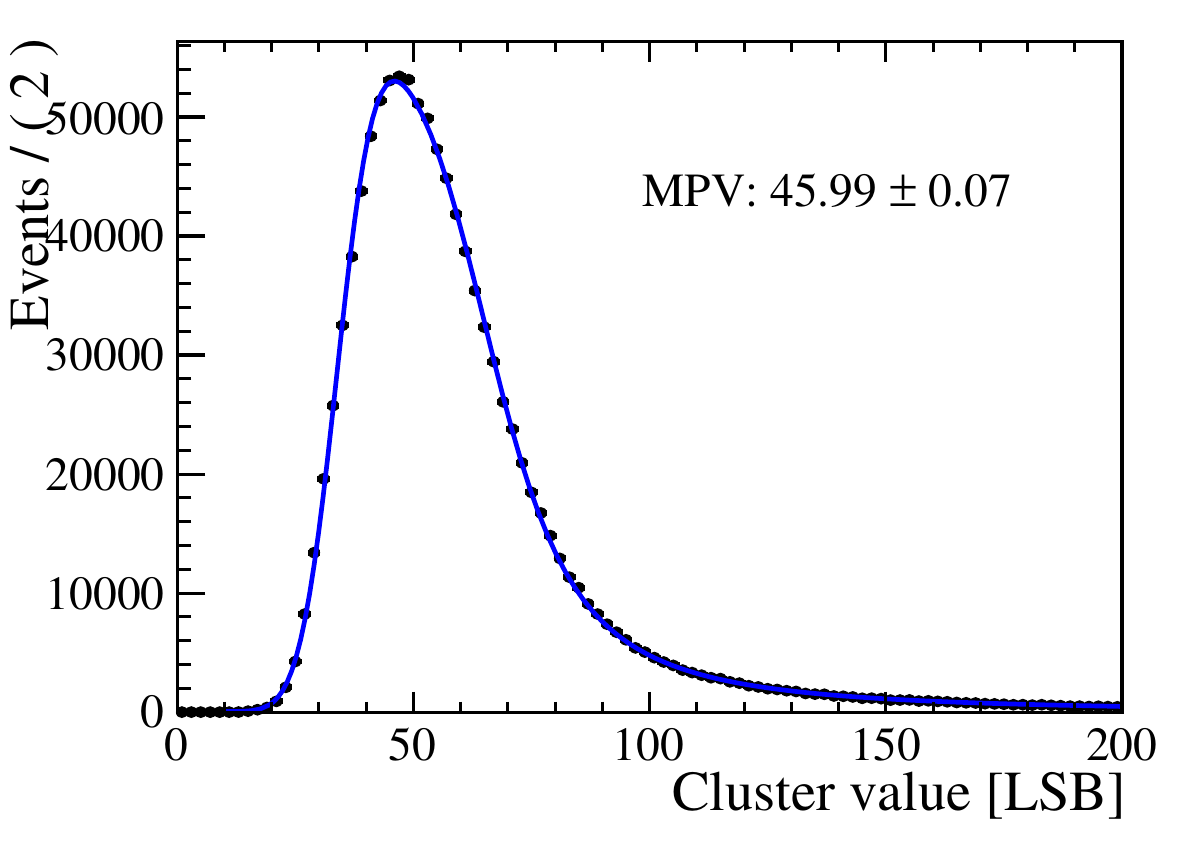}
    \end{minipage}
    }
    \subfigure[]{
    \begin{minipage}[t]{0.45\hsize}
    \centering
    \includegraphics[width=\hsize]{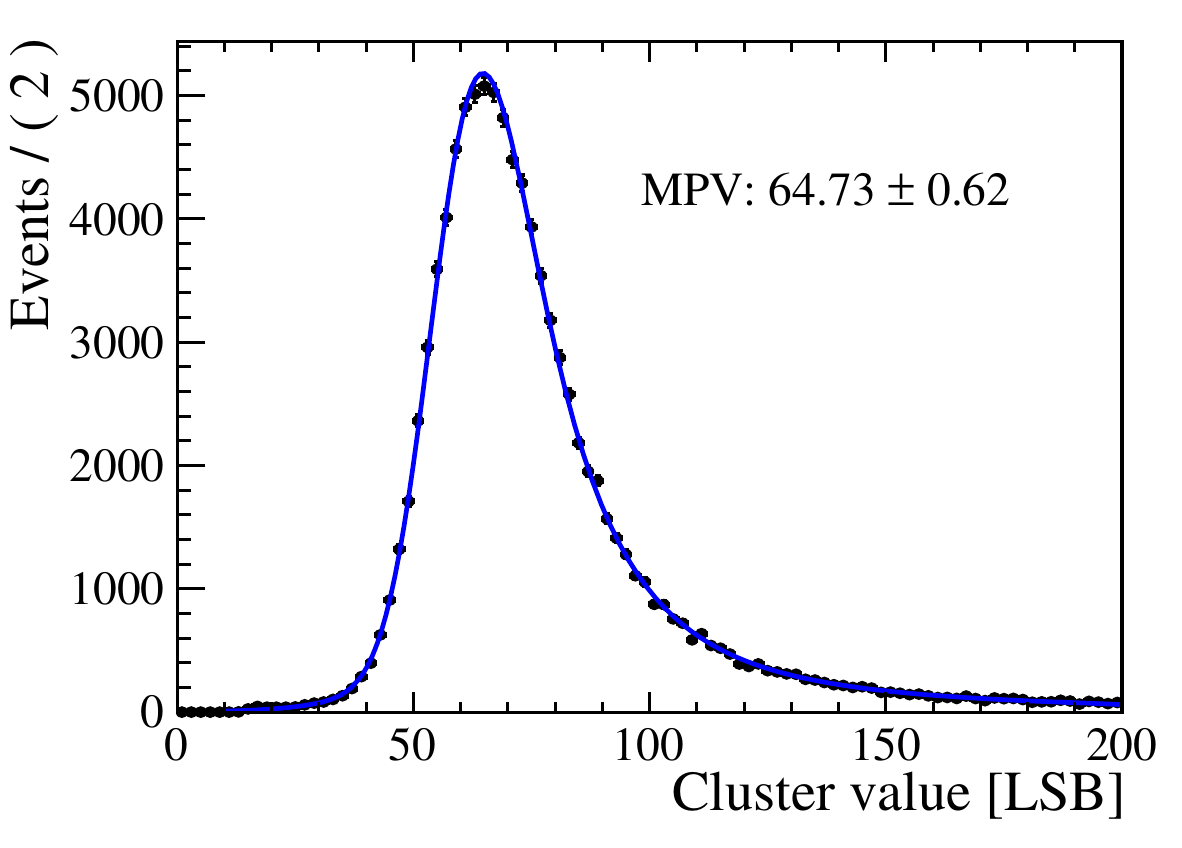}
    \end{minipage}
    }
\caption{Cluster value fitting with a Landau function convoluted with a double Gaussian function of (a) Region-A, and (b) Region-C, where MPV is indicated in the plot.}\label{fig:landau_fit}
\end{figure}

\begin{figure}[!htb] 
\centering 
\includegraphics[width=0.8\hsize]{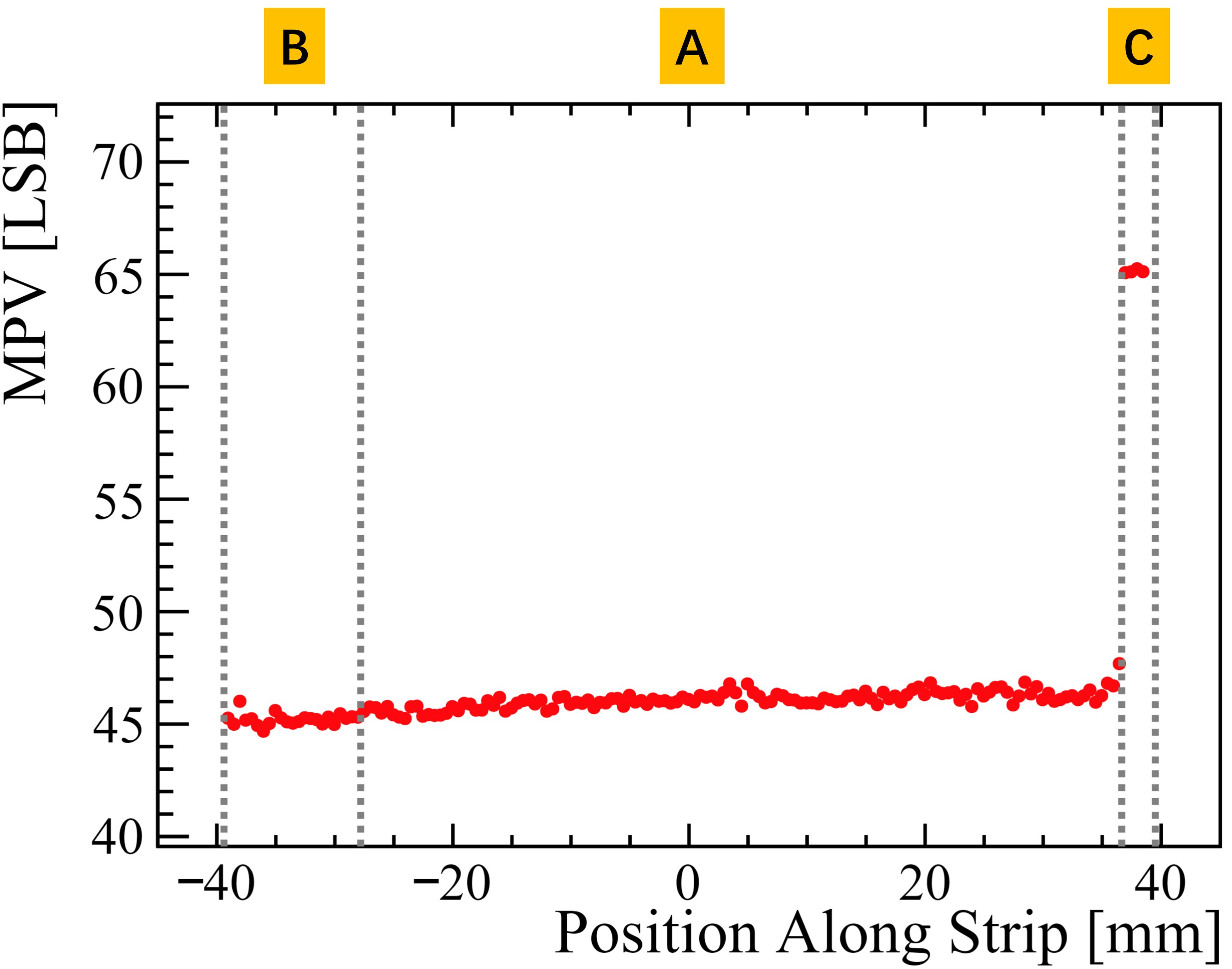}  
\caption{The MPV varies with the hit position along strip, with the dashed lines showing the SSD edge and the segmentation of the different regions as labeled.\label{fig:mpvs_along_strip}}
\end{figure}

While enhancing charge sharing, the presence of floating strips leads to a reduction in charge collection efficiency (CCE) due to capacitive coupling to backplane, signal loss through bias resistor, etc.~\cite {Casse:2013xia}. As the absolute CCE can not be determined directly, the most probable value (MPV) of a cluster signal is used to represent the relative CCE across different regions. By binning the particle incident position along strips in steps of $500 \ \mu \mathrm{m}$, the MPV was determined by fitting each bin with a Landau function convoluted with a double Gaussian function as shown in Fig.~\ref{fig:landau_fit}. And Fig.~\ref{fig:mpvs_along_strip} illustrates how the MPVs change along the strip direction. From left to right, the MPV gradually rises from $45\, \mathrm{LSB}$ to $46\, \mathrm{LSB}$, likely due to the closer proximity to the readout chip. Upon reaching Region-C, the MPV suddenly increases to $65 \, \mathrm{LSB}$, reflecting a notable improvement in CCE once the three floating strips are no longer present. While the three floating strips reduce the MPV, the SSD maintains a high detection efficiency due to the low noise level ($3 \, \mathrm{LSB}$), as detailed in Sec.~\ref{sec:efficiency}.

\subsection{Detection efficiency} \label{sec:efficiency}
In this study, the detection efficiency refers to the ability of the DUT to detect MIPs. It is one of the essential performances for silicon detectors, which is evaluated through the following procedure:
\begin{itemize}
    \item the telescope predicts a hit position \( x_{\mathrm{pred}} \) of a track on the DUT plane;
    \item the DUT then searches for cluster within a window of \( x_{\mathrm{pred}} \pm 109\, \mu\mathrm{m} \);
    \item if a cluster is found, the track is considered to have been successfully detected by the DUT;
\end{itemize}

Based on this, the efficiency along the strip is analyzed to determine the total sensitive area, and the relationship between detection efficiency and the hit position between two readout strips (hit position inner strip) is then investigated.

\begin{figure}[!htb]
\centering
\includegraphics[width=0.9\hsize]{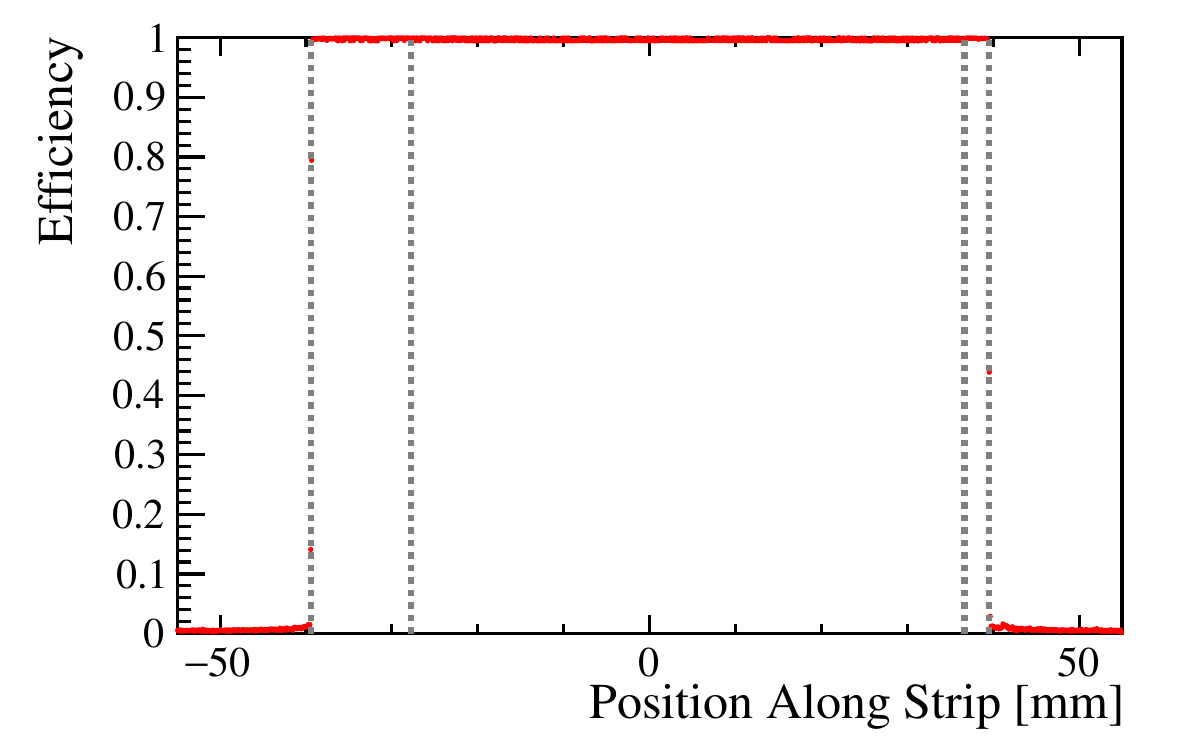}
\caption{Detection efficiency of DUT along strip direction, the dashed lines showing the SSD edge and the segmentation of the different regions.}\label{fig:eff_along}
\end{figure}

% \subsubsection{Along Strip efficiency}
Owing to the large telescope coverage, tracks spanning the full length of the SSD along the strip direction can be reconstructed. Fig.~\ref{fig:eff_along} shows the detection efficiency along the strip direction. The detection efficiency remains close to 1 across the entire SSD (total efficiency = $99.8 \, \%$), with the detector edges clearly delineated. This indicates that:
\begin{itemize}
    \item the bias resistors located between two aluminum readout strips did not affect the efficiency;
    \item the introduction of three floating strips does not lead to a reduction in detection efficiency.
\end{itemize}

\begin{figure*}
   \subfigure[]{
   \begin{minipage}[t]{0.3\hsize}
   \centering
   \includegraphics[width=\hsize]{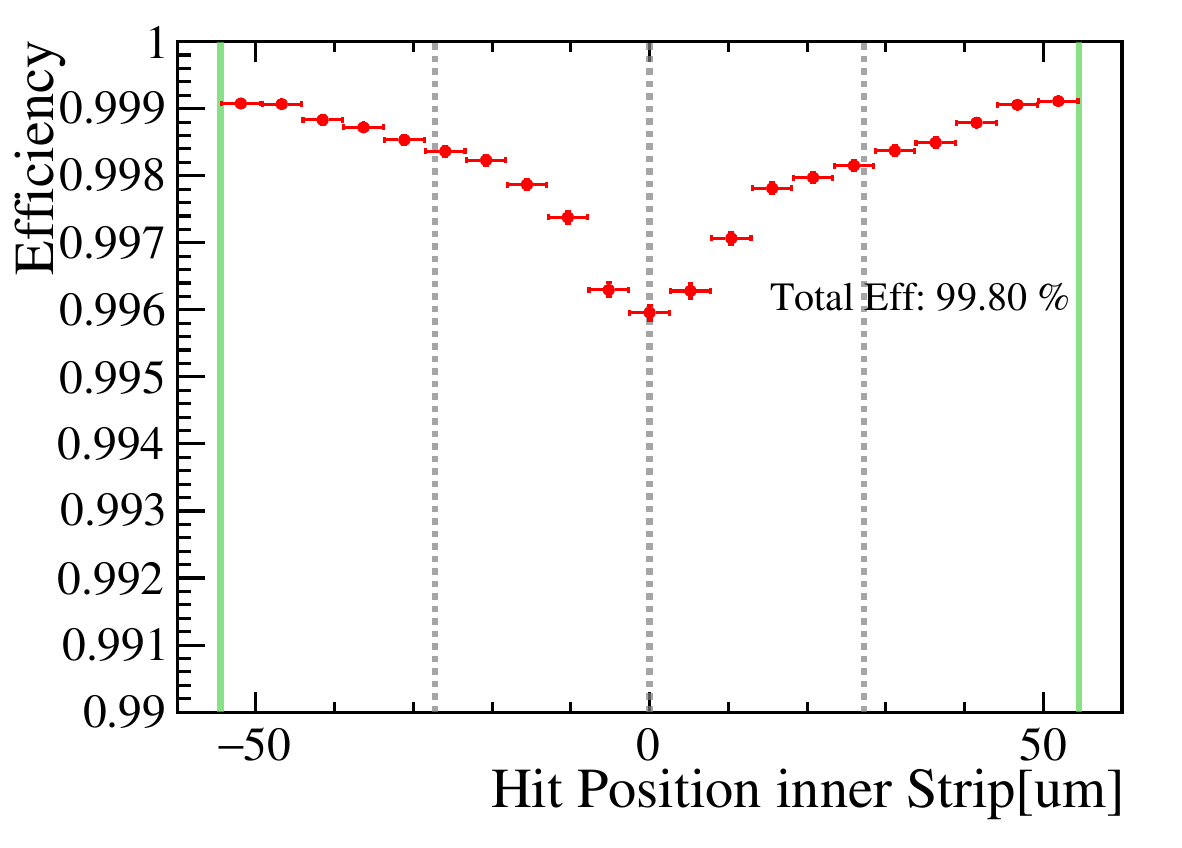}
    \end{minipage}
    }
    \subfigure[]{
    \begin{minipage}[t]{0.3\hsize}
    \centering
    \includegraphics[width=\hsize]{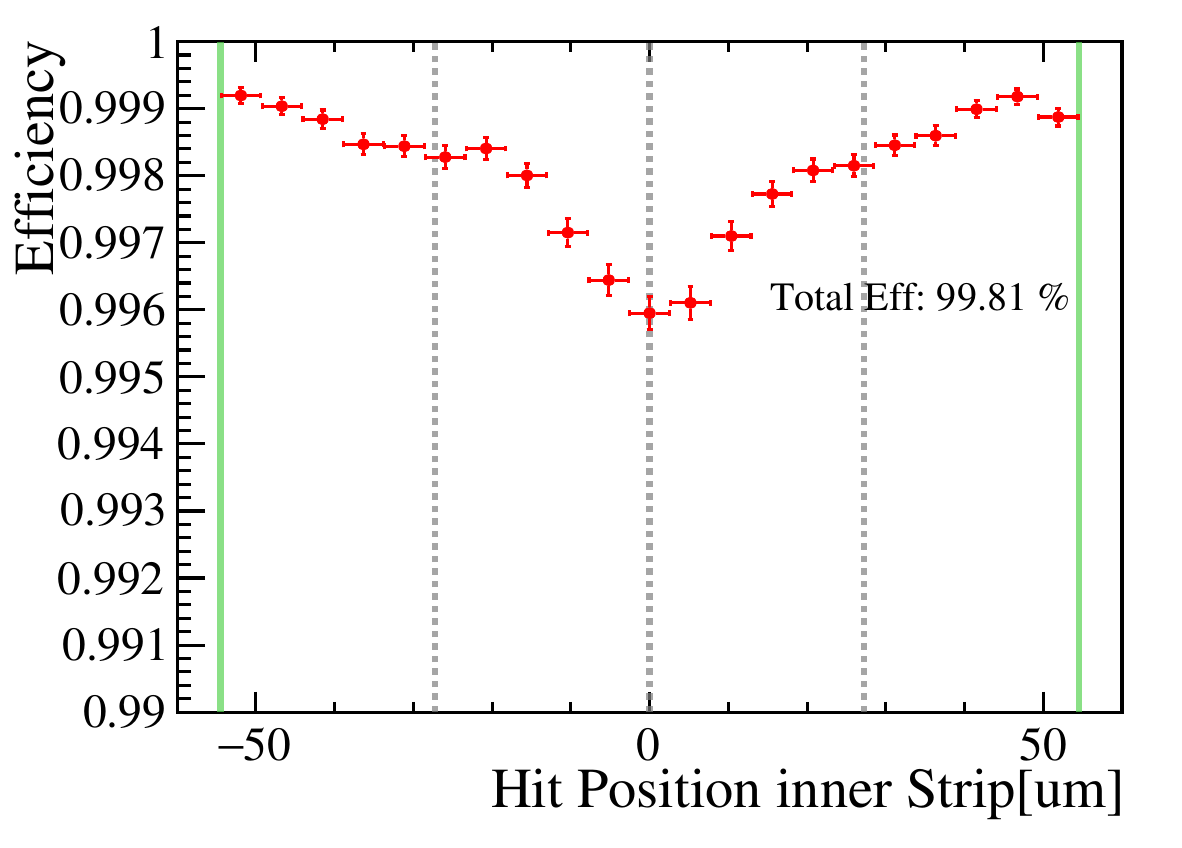}
    \end{minipage}
    }
    \subfigure[]{
    \centering
    \begin{minipage}[t]{0.3\hsize}
    \includegraphics[width=\hsize]{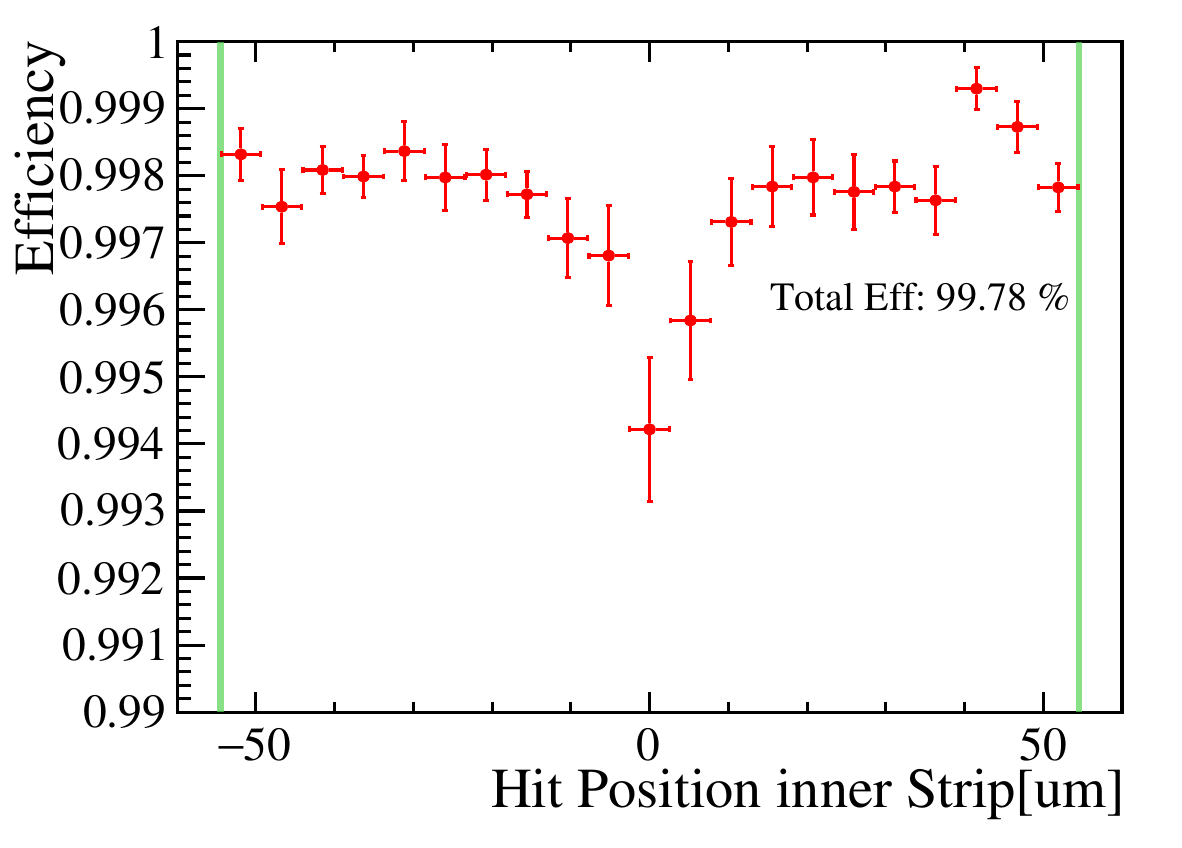}
    \end{minipage}
    }
\caption{Efficiency vs hit position inner strip of different regions: (a) Region-A, (b) Region-B, and (c) Region-C. The dashed line shows the position of the three floating strips, and the solid green lines represent the two readout strips.}
\label{fig:effic in strip}
\end{figure*}

The detection efficiency versus inner strip hit position for different regions is shown in Fig.~\ref{fig:effic in strip}. It can be seen that all regions exhibit the highest detection efficiency near the two readout strips and the lowest in the middle, which is consistent with the charge collection behavior of the SSD. For Region-A and Region-B, the efficiency decreases gradually from the readout strip to the first floating strip but drops more rapidly after crossing the first floating strip. For Region-C, with no floating strips, which results in the efficiency remaining relatively stable over a wider range, and a small drop in most central range. Generally, the total detection efficiency of all three regions are $99.8 \, \%$.

\begin{figure}[!htb]
   \subfigure[]{
   \begin{minipage}[t]{0.45\hsize}
   \centering
   \includegraphics[width=\hsize]{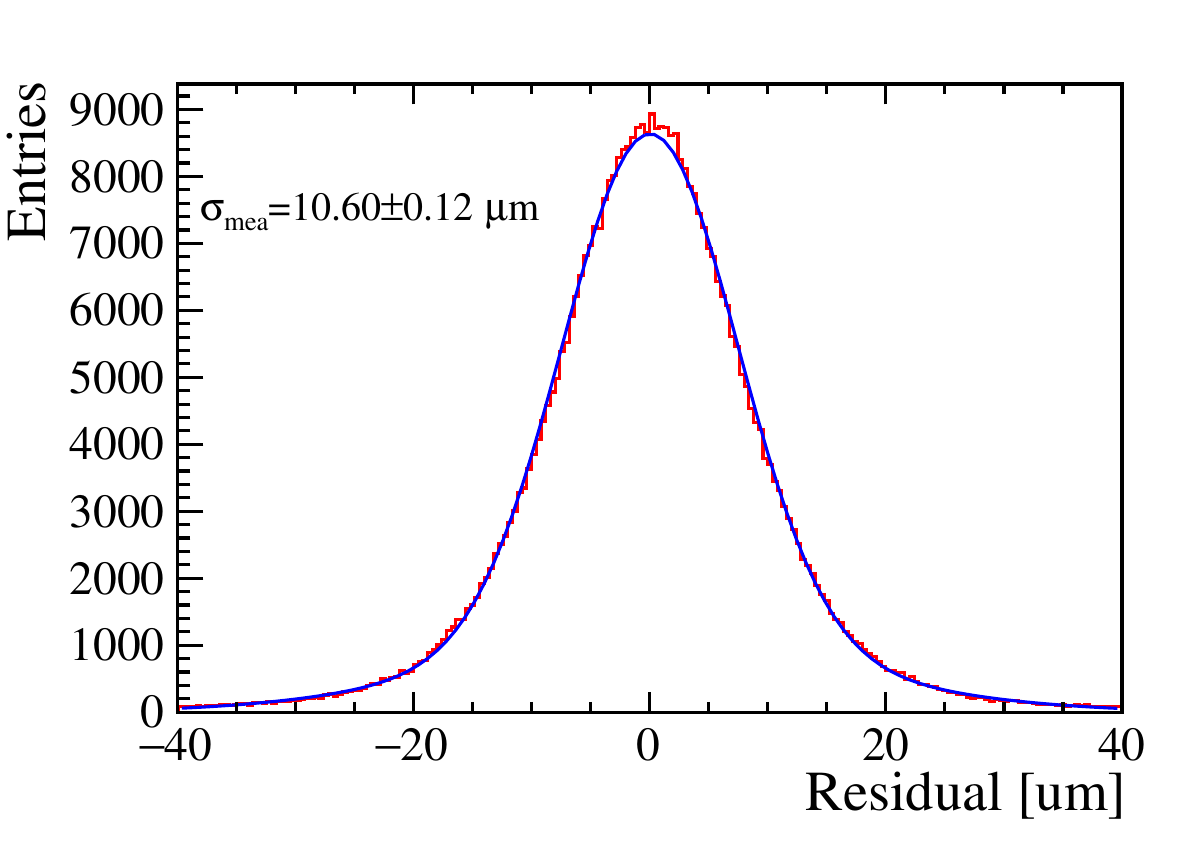}
    \label{fig:residual compare_eta1d}
    \end{minipage} 
    }
    \subfigure[]{
    \begin{minipage}[t]{0.45\hsize}
    \centering
    \includegraphics[width=1.1\hsize]{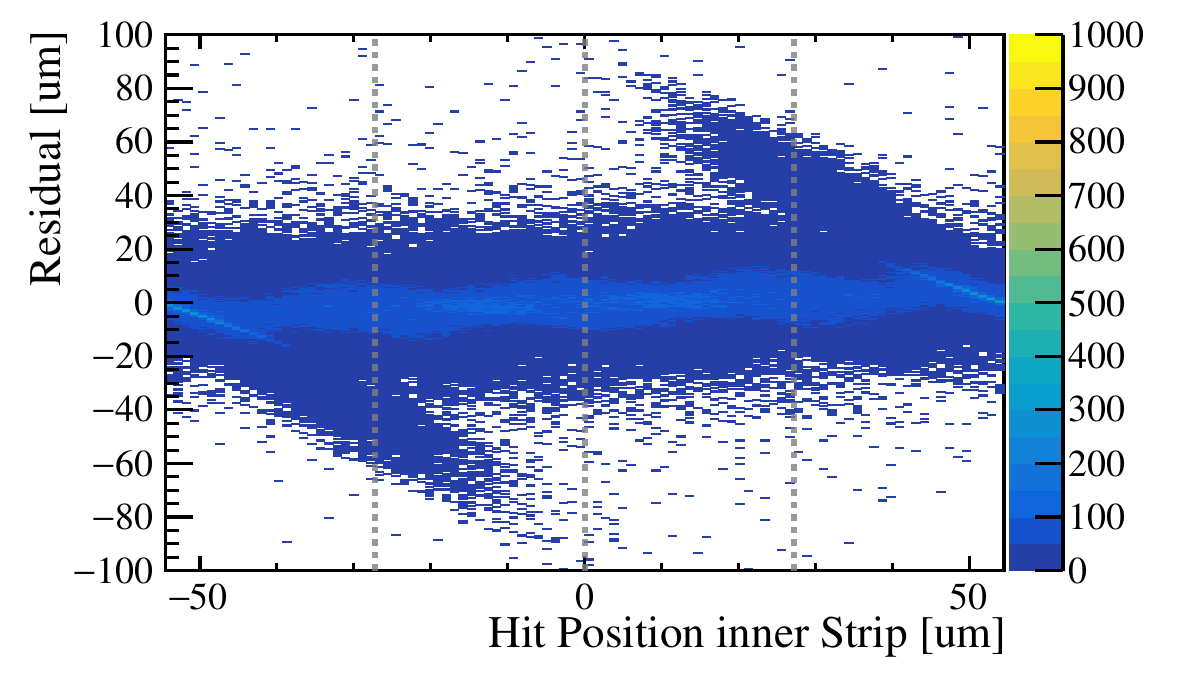}
    \label{fig:residual compare_eta2d}
    \end{minipage}
    }
    \\
       \subfigure[]{
   \begin{minipage}[t]{0.45\hsize}
   \centering
   \includegraphics[width=\hsize]{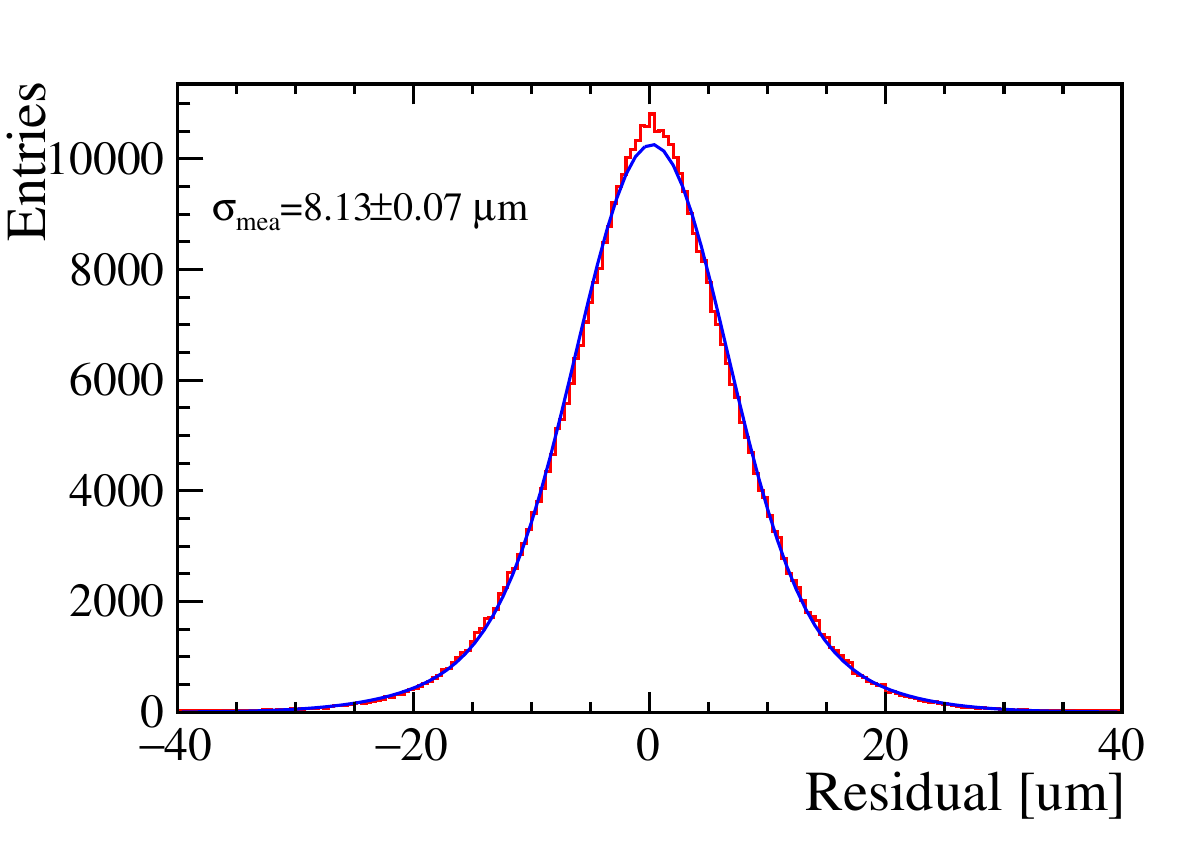}
    \label{fig:residual compare_2eta1d}
    \end{minipage}
    }
    \subfigure[]{
    \begin{minipage}[t]{0.45\hsize}
    \centering
    \includegraphics[width=1.1\hsize]{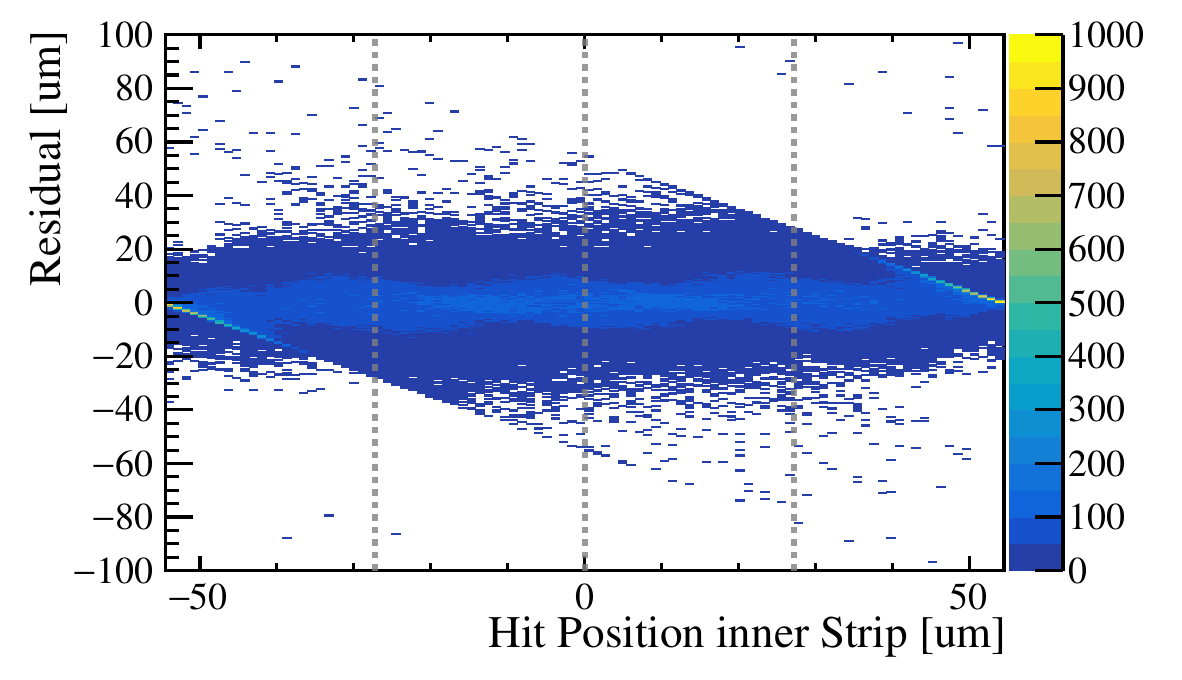}
    \label{fig:residual compare_2eta2d}
    \end{minipage}
    }
\caption{Spatial resolution comparison of ($a$)($b$) $\eta$ algorithm and ($c$)($d$) double-$\eta$ algorithm. ($a$)($c$): measured residual distribution of DUT, with $\sigma_{\mathrm{mea}}$ indicated in the plot. ($b$)($d$): measured residual vs hit position inner strip 2-D distribution, where dashed lines represent the three floating strips.
\label{fig:residual compare}}
\end{figure}

\subsection{Spatial resolution}

Spatial resolution is another key performance of silicon tracking detectors, determined by both the detector hardware and the reconstruction algorithm.  In the following, we first present the effect of the double-\(\eta\)  algorithm and then evaluate the improvements introduced by the three floating strips.

The measured unbiased resolution ($\sigma_{mea}$) is defined as the standard deviation of the difference between the position reconstructed by the DUT and the telescope prediction. It comprises two components:
\begin{equation}
    \sigma_{\mathrm{mea}} = \sqrt{\sigma_{\mathrm{dut}}^2 + \sigma_{\mathrm{tele}}^2}
\end{equation}

Where $\sigma_{\mathrm{dut}}$ is intrinsic spatial resolution of the DUT and $\sigma_{\mathrm{tele}}$ is the resolution of the telescope. Assuming each layer having the same intrinsic spatial resolution, the telescope's contribution can be readily subtracted~\cite{Li2024MCS}, giving the $\sigma_{\mathrm{dut}} = 0.89 \, \sigma_{\mathrm{mea}}$ in our beam test setup.

\subsubsection{Effect of the double-\(\eta\) Algorithm \label{sec:compare_double_eta}}

The measured residual distributions for Region-A obtained using the \(\eta\) algorithm and the double-\(\eta\) algorithm are shown in Fig.~\ref{fig:residual compare}(a)(a), respectively. It can be seen that the first one exhibits a broader distribution, with pronounced tails on both sides. Meanwhile, in the two-dimensional distribution of measured residual versus inner strip hit position shown in Fig.~\ref{fig:residual compare}(c), a clear Z-shape is observed, with the two horns contributing to the tails in Fig.~\ref{fig:residual compare}(a). These events correspond to the ``swapping phenomenon" discussed in Sec~\ref{sec:doubel_eta}. After applying the correction using the double-\(\eta\) algorithm, the Z-shaped structure disappears, as shown in Fig.~\ref{fig:residual compare}(d). As a result, the $\sigma_{\mathrm{mea}}$ was optimized from $10.6 \, \mu \mathrm{m}$ to $8.1 \, \mu \mathrm{m}$, and the intrinsic $\sigma_{\mathrm{dut}}$ improves from $9.4 \, \mu \mathrm{m}$ to $7.2 \, \mu \mathrm{m}$. All the spatial resolution results discussed below are obtained using the double-$\eta$ algorithm.

\begin{figure*}
   \subfigure[]{
   \begin{minipage}[t]{0.3\hsize}
   \centering
   \includegraphics[width=\hsize]{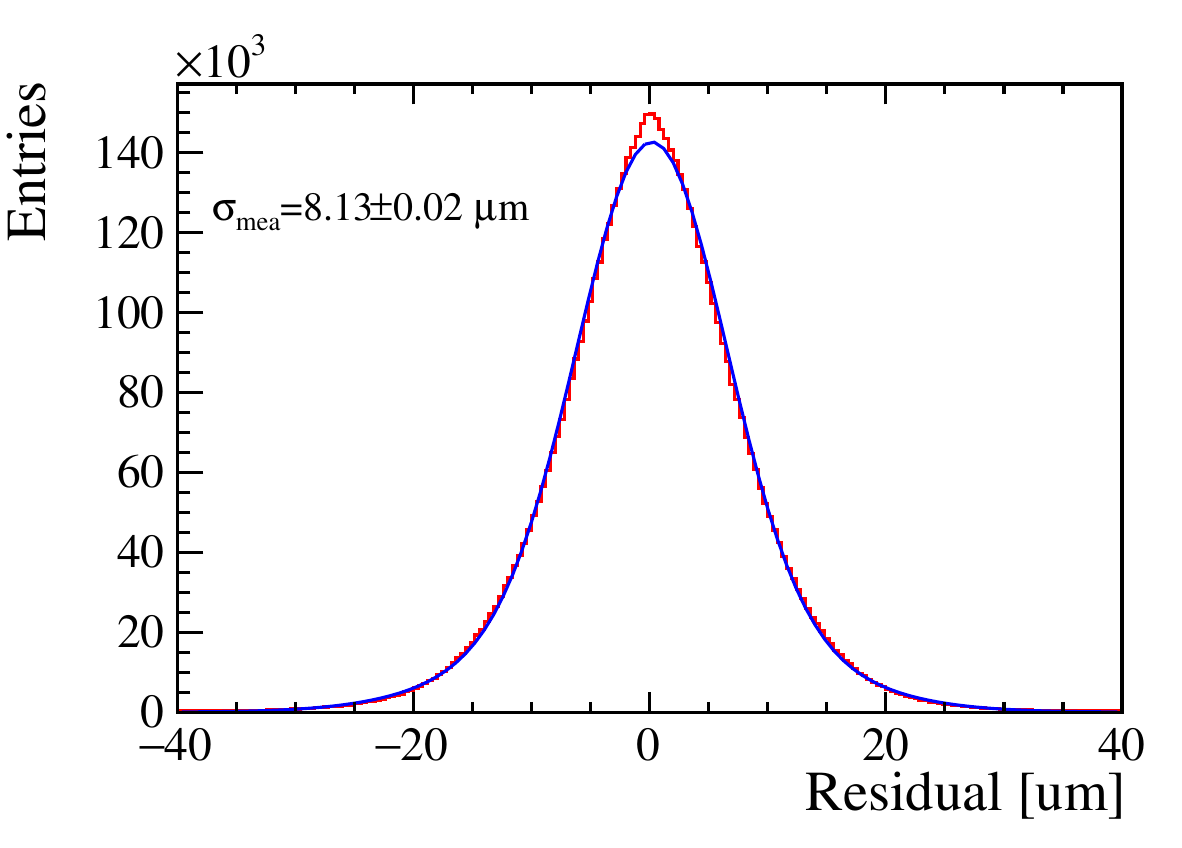}
\label{fig:residual compare region_A}
    \end{minipage}
    }
    \subfigure[]{
    \begin{minipage}[t]{0.3\hsize}
    \centering
    \includegraphics[width=\hsize]{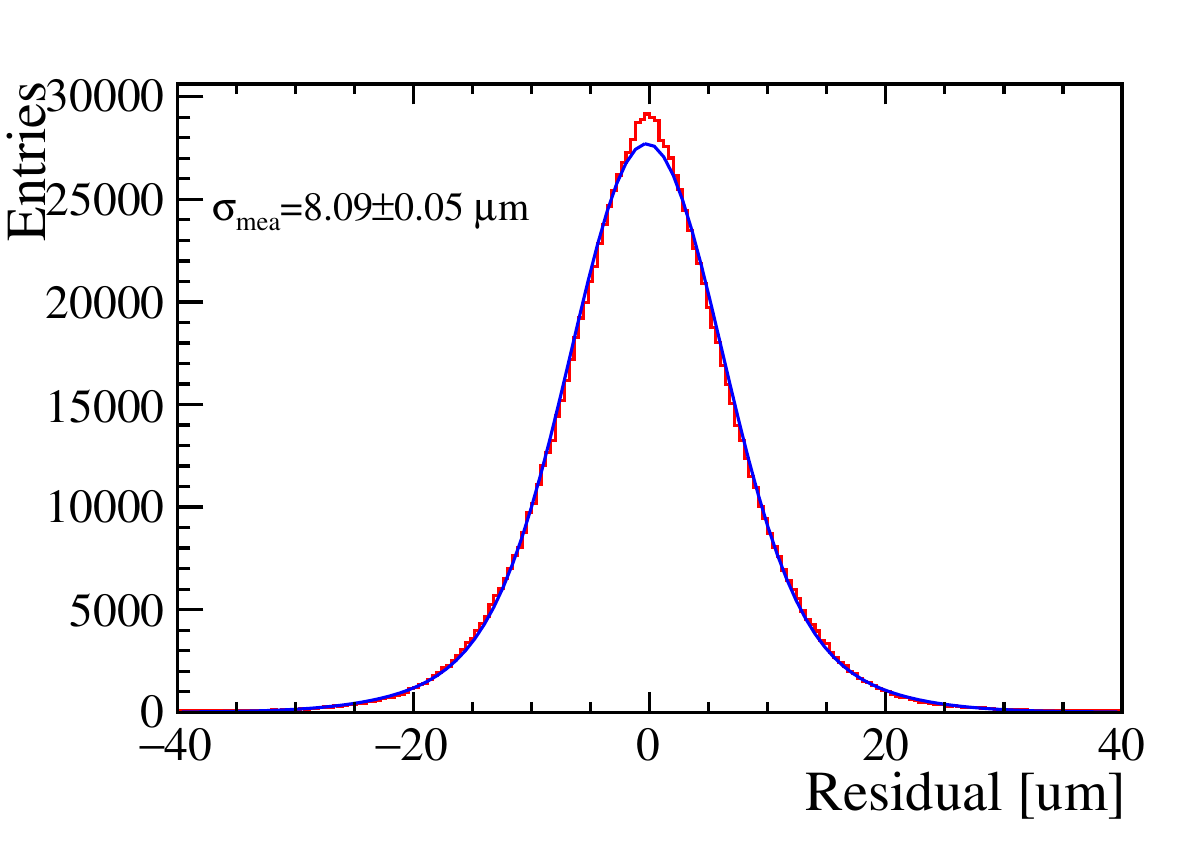}
\label{fig:residual compare region_B}
    \end{minipage}
    }
    \subfigure[]{
    \centering
    \begin{minipage}[t]{0.3\hsize}
    \includegraphics[width=\hsize]{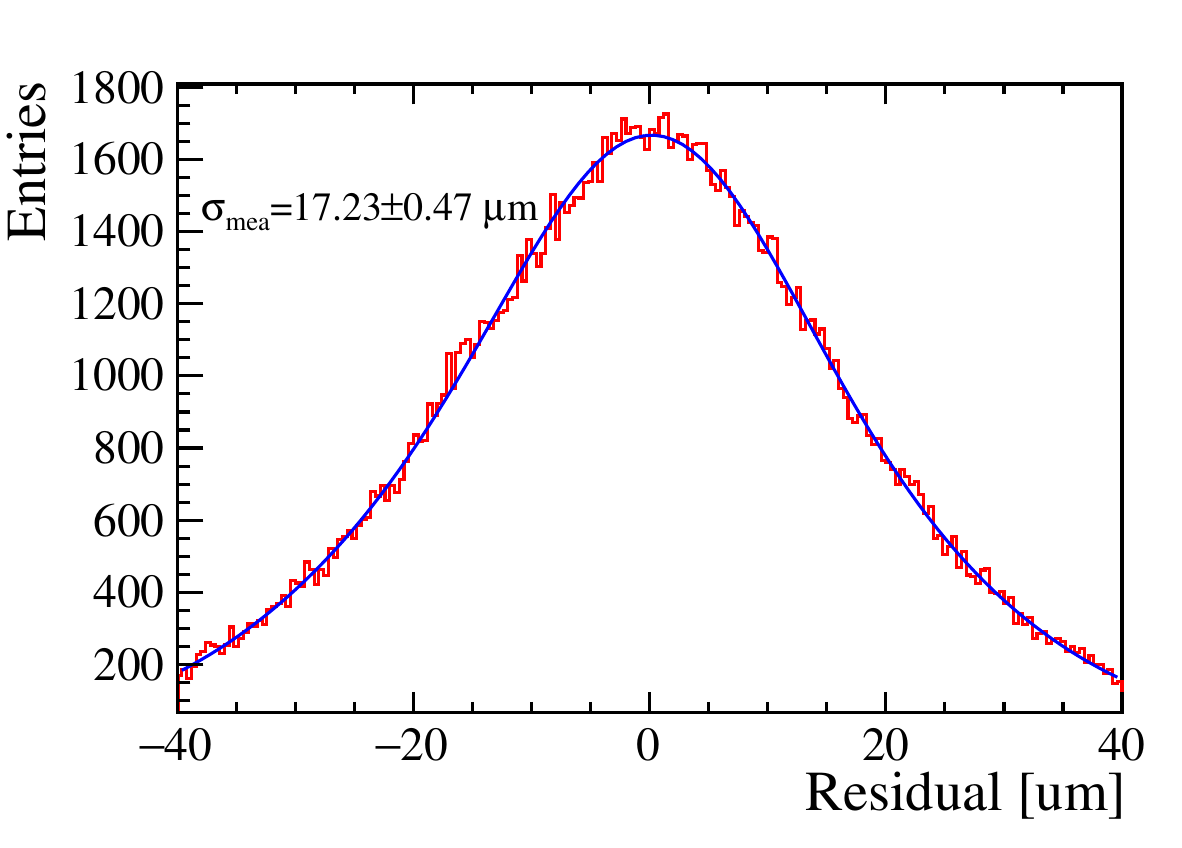}
\label{fig:residual compare region_C}
    \end{minipage}
    }
\caption{Measured residual distribution of different regions :(a) Region-A, (b) Region-B, and (c) Region-C. The $\sigma_{\mathrm{mea}}$ indicated in the plot.
\label{fig:residual compare region}}
\end{figure*}

\subsubsection{Improvement from the three floating strips}

The measured residual distribution with and without the three floating strips as shown in Fig.\ref{fig:residual compare region}(a)(c). The three floating strips significantly improves the $\sigma_{\mathrm{mea}}$ from $17.2 \, \mu \mathrm{m}$ to $8.1 \, \mu \mathrm{m}$ ($\sigma_{\mathrm{dut}}$ from $15.2 \, \mu \mathrm{m}$ to $7.2 \, \mu \mathrm{m}$). This enhancement is primarily due to the larger charge sharing effect introduced by the floating strips. This can also be demonstrated by comparing the $\eta$ distributions in the two regions as shown in Fig.~\ref{fig:eta compare}(a)(c). The three floating strips lead to a flatter \(\eta\) distribution, indicating increased sensitivity of the \(\eta\) value to the inner strip hit position. As a result, the function \( f(\eta) \) becomes closer to the linear function ($f(\eta) = \eta$), as shown in Fig.~\ref{fig:eta compare}(b)(d), which leads to better spatial resolution~\cite{Turchetta:1993vu}. In addition, Fig.~\ref{fig:residual compare region}(b) shows the same plot of the residual in region-B. The $\sigma_{\mathrm{mea}}$ ($\sigma_{\mathrm{dut}}$) of this region is $8.1 \, \mu \mathrm{m}$ ($7.2 \, \mu \mathrm{m}$), indicating that placing bias resistors between two aluminum readout strips does not affect the spatial resolution. 

Eventually, by area-weighting the three regions, the overall spatial resolution of the SSD is $7.6 \,\mu\mathrm{m}$ for MIPs.

\section{Summary}
We have designed an SSD for large-area, power-limited applications and did a detailed performance characterization for a single sensor by beam test. Introducing three floating strips and inserting their bias resistors between two aluminum readout strips are the distinctive designs of this SSD. Our study shows that the three floating strips significantly enhance the charge sharing effect, which improved spatial resolution from $15.2 \, \mu \mathrm{m}$ to $7.2 \, \mu \mathrm{m}$ without compromising detection efficiency. And the presence of bias resistors between two aluminum strips does not degrade the spatial resolution or detection efficiency in the corresponding region. As a result, the SSD gets an overall efficiency of $99.8 \, \%$ and spatial resolution $7.6 \, \mu \mathrm{m}$ for MIPs. During this study, a double-\(\eta\) algorithm for hit position reconstruction is developed to suit this SSD. The SSD from this design has been successfully applied in the AMS Layer-0 tracker upgrade, and the design principles can be utilized for future silicon detectors in space-borne cosmic ray experiments.
\\

\section*{Acknowledgments}
This study was supported by the National Key Programme for S\&T Research and Development (Grant NO.: 2022YFA1604800), and the National Natural Science Foundation of China (Grant NO.: 12342503). We express our gratitude to our colleagues in the CERN accelerator departments for the excellent performance of the SPS.

\bibliographystyle{unsrt}  
\bibliography{nst_template}    

\end{document}